# MAIEI

# THE MAIEI LEARNING COMMUNITY REPORT

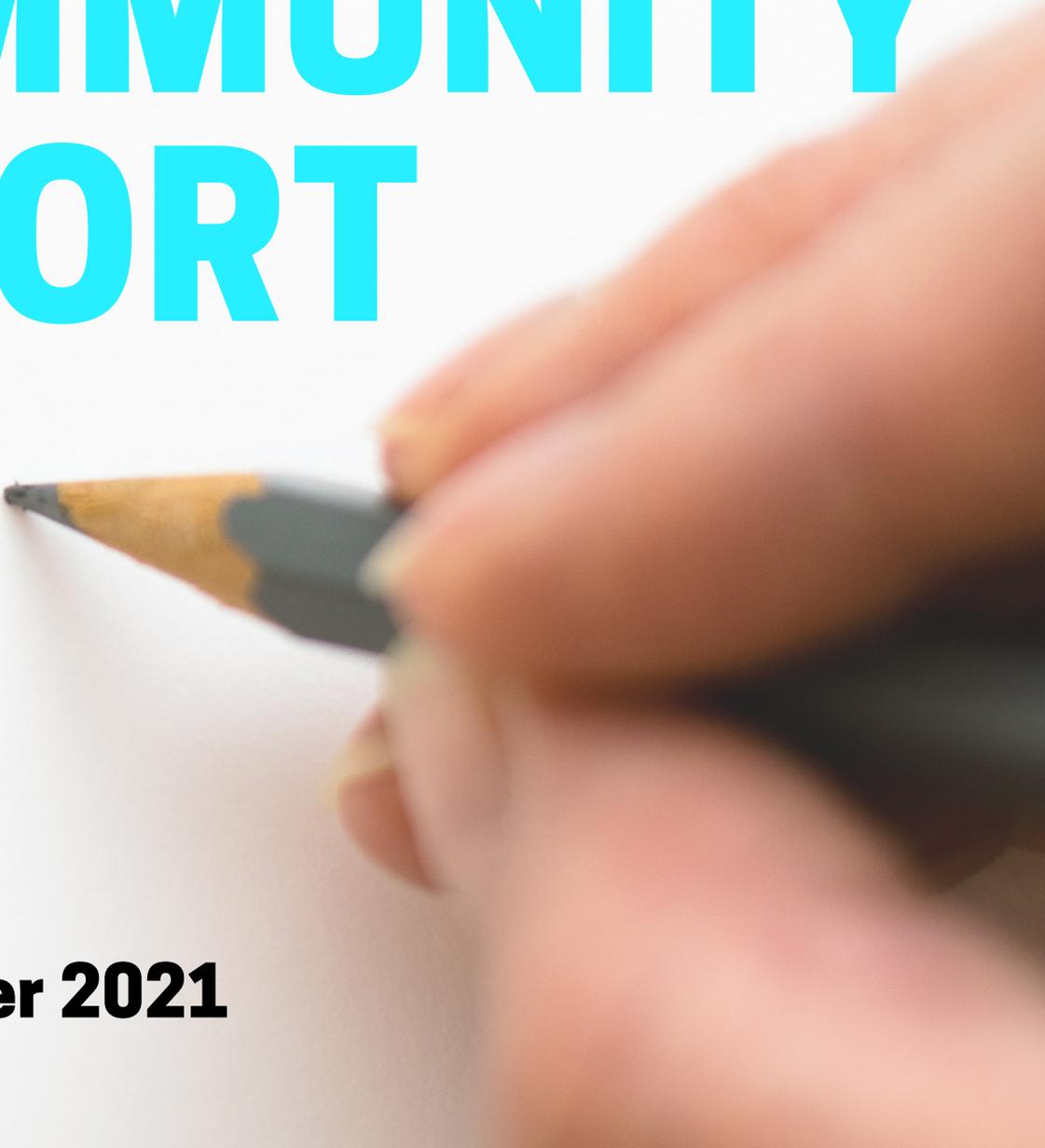

**September 2021**

This report was compiled by the team at the Montreal AI Ethics Institute (MAIEI) as a result of our 1st Learning Community Cohort. Learn more about how we are democratizing AI literacy on our website and subscribe to our weekly newsletter The AI Ethics Brief!

This work is licensed open-access under a Creative Commons Attribution 4.0 International License.

Primary contact for the report: **Abhishek Gupta** (abhishek@montrealethics.ai)

## The Team at MAIEI

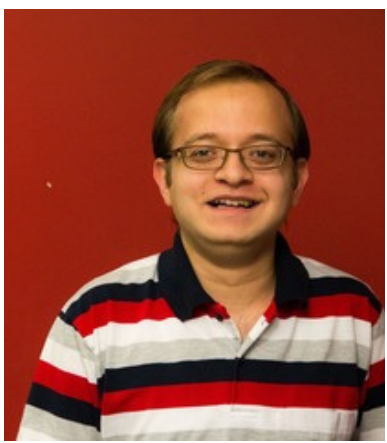

**Abhishek Gupta**
Founder, Principal Researcher and Director

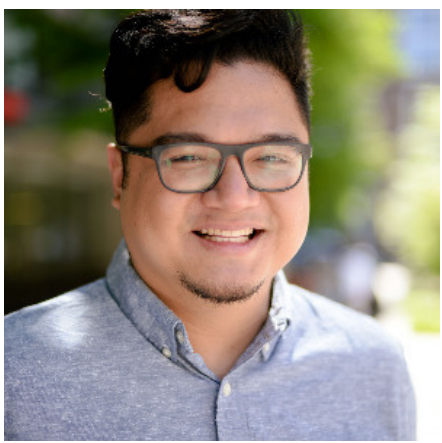

**Renjie Butalid**
Co-founder and Director

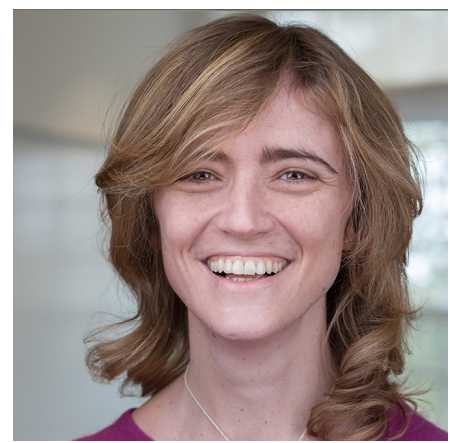

**Marianna Ganapini, PhD**
Faculty Director

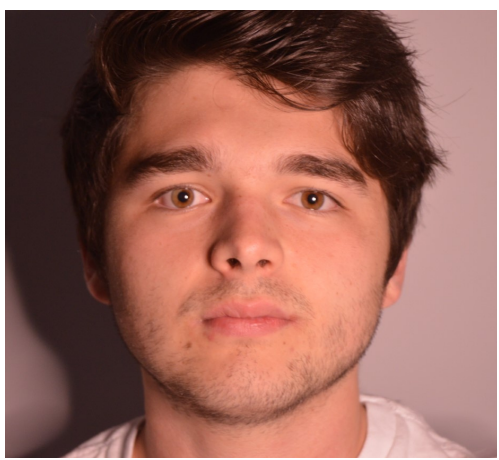

**Connor Wright**
Partnerships Manager

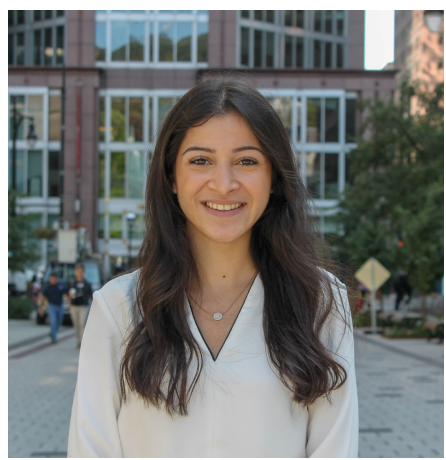

**Masa Sweidan**
Business Development Manager

The AI Ethics Brief   |   The MAIEI Learning Community Report

# FACILITATORS

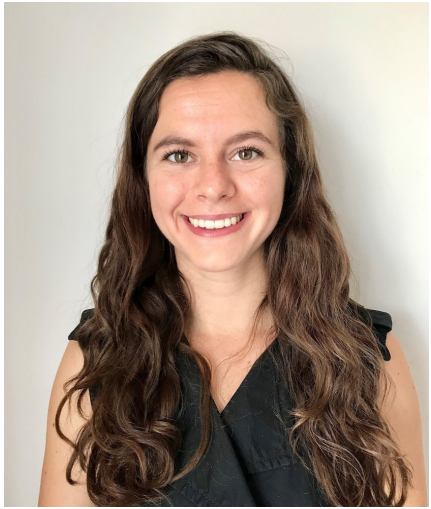

**Alexandrine Royer**
Former Educational Programme Manager

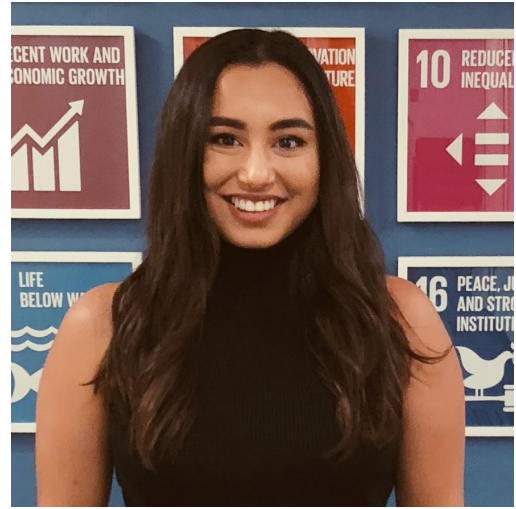

**Muriam Fancy**
Former Network Engagement Manager

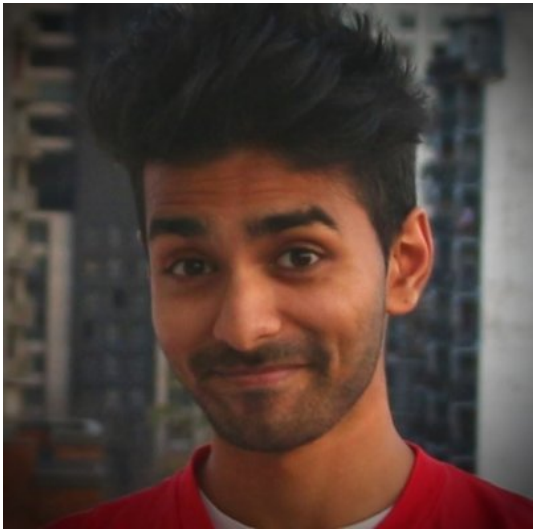

**Mo Akif**
Former Director of Communications

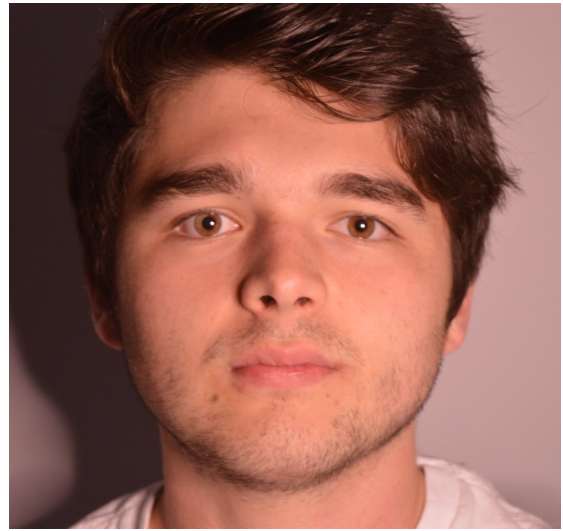

**Connor Wright**
Partnerships Manager

# LEARNING COMMUNITY COHORT

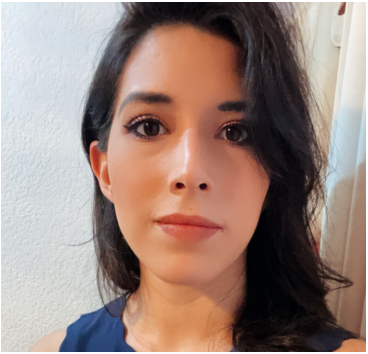

**Victoria Martín del Campo**
Department Chief, Mexico Data Strategy and Digital transformation office.

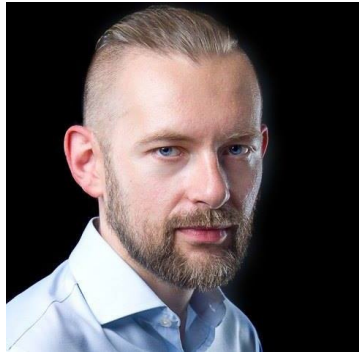

**Yurii Haidai**
Investigator, Business Ombudsman of Ukraine.

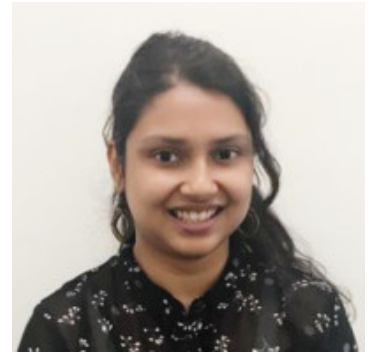

**Nanditha Narayanamoorthy**
Fourth-year doctoral candidate, Department of Humanities, York University, Toronto, Canada

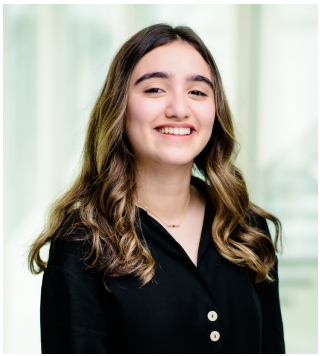

**Lujain Ibrahim**
Schwarzman Scholar, Master's in Global Affairs, Tsinghua University, Beijing, China.

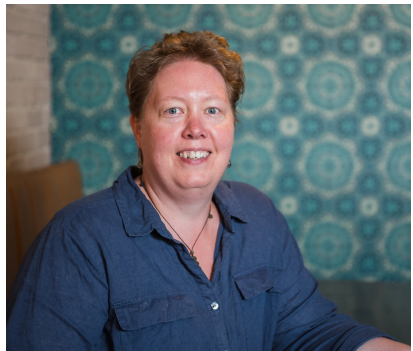

**Heather von Stackelberg**
Writer, instructional designer and educator, intersection of science, technology and society.

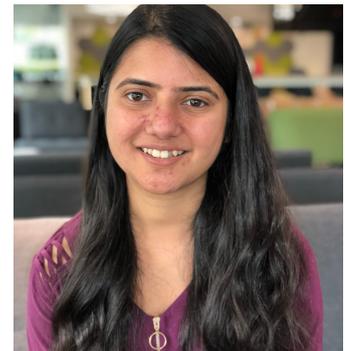

**Shreyasha Paudel**
independent researcher, ethical and social impact of automation, digitization and data in Nepal

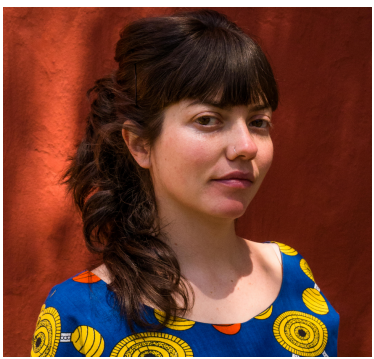

**Sofia Trejo**
Professor, Department of Mathematics, Mexico's Autonomous Institute of Technology (ITAM)

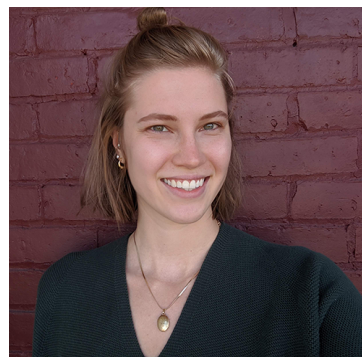

**Christina Isaicu**
Management team, AI4Good Lab

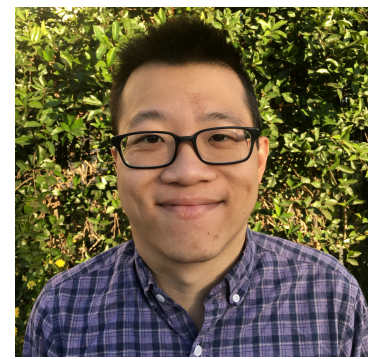

**Wilson Lee**
Senior Machine Learning Research Engineer at The Trevor Project

# LEARNING COMMUNITY COHORT

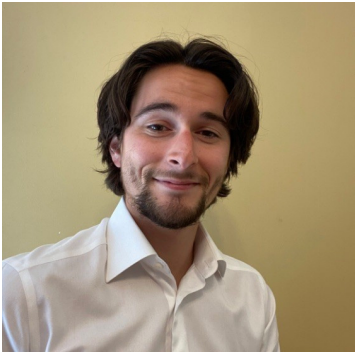

**Mitchel Fleming**
Graduate; McGill Faculty of Law, articling student with Bereskin & Parr Toronto.

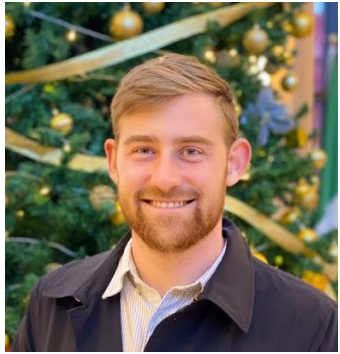

**Samuel Curtis**
AI Policy Researcher & Project Manager at The Future Society.

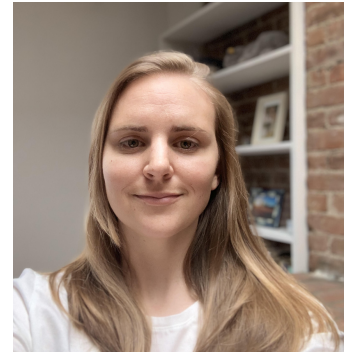

**Brittany Wills**
Software engineer, Twitter's Machine Learning Ethics, Transparency, and Accountability team.

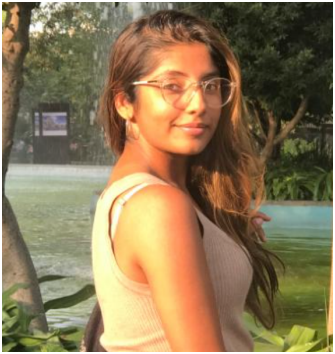

**Garima Batra**
Software engineer and designer, Master's in Information science University of Toronto.

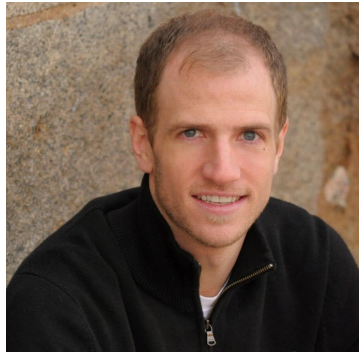

**Matthew Hutson**
Freelance science and technology writer, Contributing Writer, The New Yorker.

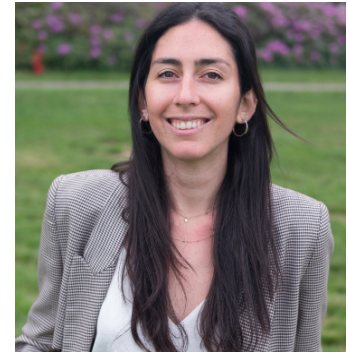

**Tiziana Zevallos**
Design Researcher at All In.

# TABLE OF CONTENTS

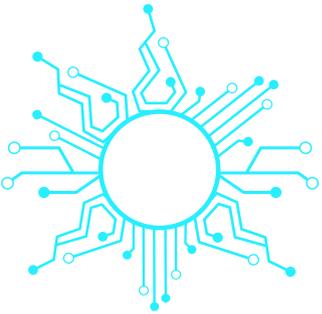

# 01

# A NOTE FROM OUR FOUNDER

Welcome to this special report from the Montreal AI Ethics Institute!

This is a labor of the Learning Community cohort that was convened by MAIEI in Winter 2021 to work through and discuss important research issues in the field of AI ethics from a multidisciplinary lens. The community came together supported by facilitators from the MAIEI staff to vigorously debate and explore the nuances of issues like bias, privacy, disinformation, accountability, and more especially examining them from the perspective of industry, civil society, academia, and government.

The outcome of these discussions is reflected in the report that you are reading now - an exploration of a variety of issues with deep-dive, critical commentary on what has been done, what worked and what didn't, and what remains to be done so that we can meaningfully move forward in addressing the societal challenges posed by the deployment of AI systems.

The chapters titled "Design and Techno-isolationism", "Facebook and the Digital Divide: Perspectives from Myanmar, Mexico, and India", "Future of Work", and "Media & Communications & Ethical Foresight" will hopefully provide with you novel lenses to explore this domain beyond the usual tropes that are covered in the domain of AI ethics.

The cohort was an absolute delight and the interactions of both the staff and the community brought together a unique, irreplaceable experience that is reflected in the following pages. If you'd like to learn more about such programs, I encourage you to sign up for The AI Ethics Brief.

Enjoy these pages and please do stay in touch!

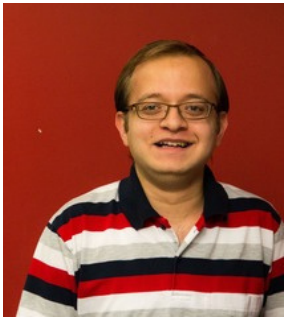

Abhishek Gupta
Founder and Principal Researcher, Montreal AI Ethics Institute
Machine Learning Engineer and CSE Responsible AI Board Member, Microsoft
Chair, Standards Working Group, Green Software Foundation
https://atg-abhishek.github.io

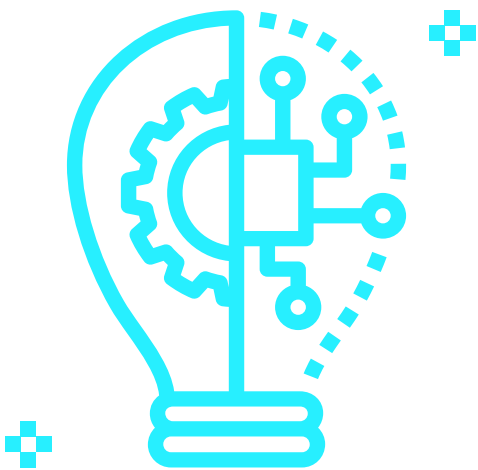

The AI Ethics Brief   |   The MAIEI Learning Community Report

# 02 INTRODUCTION

From its inception, the MAIEI LC community sought to combine both learning and communication. Learning was taken in the form of a wide range of topics over the 8-weeks, covering topics like the rise of emotional systems all the way to digital labor. Communication then came in the form of LC participants sharing their thoughts with fellow participants and coordinators. What made this even more noteworthy was shear interdisciplinarity of the group.

With a wealth of talented applications to choose from in the beginning, the team at MAIEI knew this was going to be an intellectually stimulating and riveting 8-weeks. The backgrounds included ranged from political, to legal and minority expertise. With our participants also spread all over the world, the wealth of diversity in experience and how beneficial this can be to the AI process, was extremely apparent. Whether it be talking about connectivity issues in Mexico or recent issues in Ukraine, we were all able to learn about topics previously unknown to some of us. It was with great sadness that the shared learning space came to an end, but we couldn't just leave what we learnt confined to the Zoom room. Thus, the MAIEI LC report was born.

Consisting in four chapters, the various teams and individuals have collaborated to immortalize their learnings and convey what they have discovered. Each chapter is different both in style and in content, which owes to the different experiences and perspectives present in the report. The journey for the teams to incorporate these views was long and at times clouded, but the beacon that is communication was utilised to full effect and with utter determination by all. It is a testament to every members' hard work that the report is now a reality.

We are extremely proud of what the group has achieved, and we are excited for you to see why.

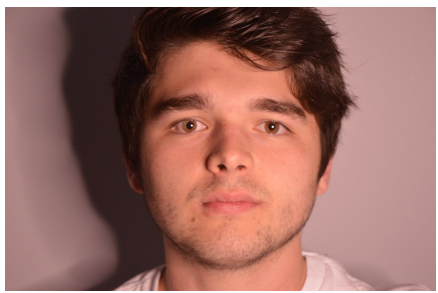

**Connor Wright, Partnerships Manager**

[Connor Wright](#)

[@Csi_wright](#)

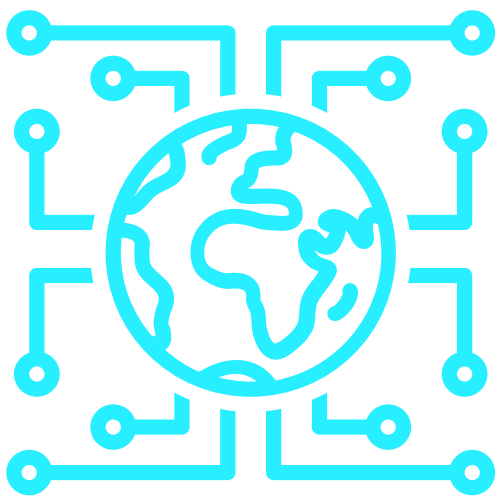

The AI Ethics Brief   |   The MAIEI Learning Community Report

# DESIGN AND TECHNO-ISOLATIONISM

**Mitchel Fleming, Brittany Wills, Christina Isaicu, Wilson Lee, Tiziana Zevallos.**

Introduction

The format of this chapter is intended to reflect and extend the spirit of the [MAIEI Learning Community](#) by documenting a conversation between five people who bring differing educational, experiential, and industry perspectives to the discussion about design and techno-isolationism in AI.

The chapter will open with a brief introduction about the five authors in an effort to make apparent the perspectives that will inform the discussion.

In undertaking such a conversation, we hope to enact the thesis of our argument; interdisciplinary collaboration and dialogue between diverse perspectives fosters potential for growth. In having this conversation, we can also directly highlight the limitations in our own perspectives ("I feel I cannot talk about X because I lack Y"). The uniqueness and limitations of any given perspective go hand-in-hand, so the goal is not *simply* to collect diverse perspectives; by encouraging *interaction* between these perspectives we, as active participants in the conversation, seek to evolve with it.

The conversation will follow an abridged journey map of how AI systems are currently being designed and developed. By identifying the different actors and domains involved in the process of developing AI, we will illustrate the need for interdisciplinary collaboration in AI by highlighting the limitations (and consequences) encountered when each "step" occurs in isolation. We also wish to express that this exercise is in and of itself isolated from the wider context in which AI is developed and implemented. From the mining of minerals, to the user experience, to waste disposal, one cannot take into account the true scope of AI's global impact without looking at the entire supply chain. That said, for the purposes of this publication, we will be focusing exclusively on the software development process.

Throughout the discussion, it will become apparent that even as a group of five individuals with differing perspectives, we are still limited in our ability to fully encapsulate the scope of this conversation.
By being transparent about our own limitations, we invite the opportunity for further collaboration and understanding and reflect the ideals we propose as we discuss what interdisciplinary AI could look like.

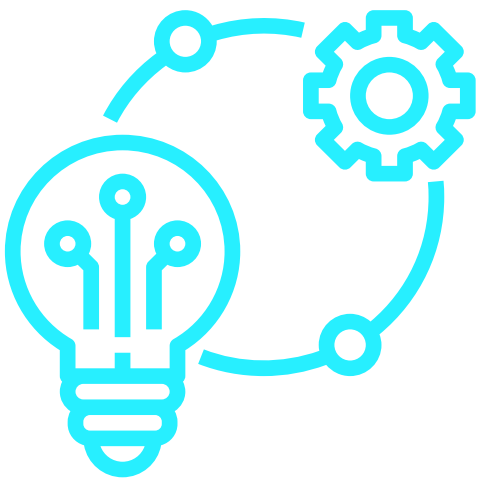

## Meet the members

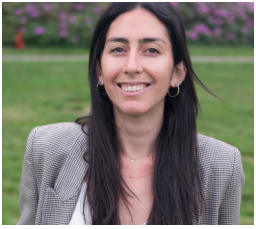

**Tiziana Zevallos** is a Design Researcher at All In, a Human-Centred Design agency in New Brunswick, Canada, that partners with non-profits and the public sector to improve their services.

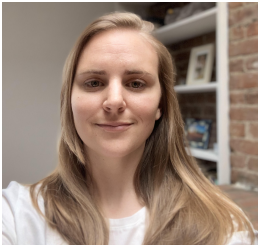

**Brittany Wills** is a software engineer on Twitter's Machine Learning Ethics, Transparency, and Accountability team.

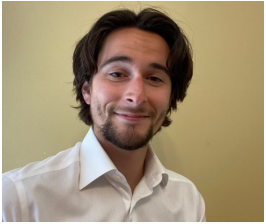

**Mitchel Fleming** is a former Director and CEO at Halucenex, a life sciences company developing psychedelic compounds for mental health care. He is also a soon-to-be articling student at Bereskin & Parr, an award-winning IP law firm.

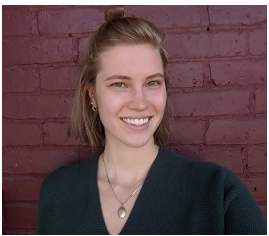

**Christina Isaicu** runs operations and marketing at the AI4Good Lab in Montreal, a non-profit AI education program that provides training and mentorship to women.

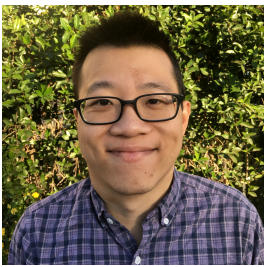

**Wilson Lee** conducts research in Natural Language Processing at The Trevor Project, a non-profit organization that offers crisis intervention and suicide prevention services to LGBTQ youth.

### Christina Isaicu

"Towards the end of the Learning Community sessions, we had all been brainstorming how we were going to execute this conversation. At one point in the conversation, I had stated that the development of AI must certainly be a very interdisciplinary process given the fact that it is a tool that is applied across various industries in various contexts. I thought that the sheer multitude of actors and individuals affected would demand collaboration! And as someone who works in an education-focused, not-for-profit diversity space, I carry a specific worldview on how these projects should occur. And my assertion ended up being wrong, as Brittany went on to explain! She explained to me that the whole process, in practice, is much more disjointed than what one might hope for or expect. Different areas of the development process seldom interact with or understand the activities performed by prior or subsequent developers, leading to techno-isolationism throughout the development process. This was critical in informing my understanding of how these processes occur, thereby fundamentally altering my perception of reality.

So it was from that interaction that we set about having this conversation: to explore and contrast the ideations of some, and the realities of others, to bridge the gap between what is and what could or should be. We hoped that in having this conversation, we might be able to understand why things are the way they are and how we might be able to proceed to a world where multidisciplinary collaboration in AI is the default rule. We hope that this conversation can act as a template for others working in this industry to bridge the gap and foster meaningful dialogue."

How it is & How it could be

Framing the Problem

When tasked to solve a problem, teams will shoot for the stars and aim to develop a Machine Learning (ML) solution that is at least on par with state of the art solutions. Through our conversation, we called this the two tribes dilemma: Product Problem vs Research Problem. The Product Problem refers to when the team looks at a problem and its success is measured through a defined metrics lens. The Research Problem refers to the technical research aspect of ML and asks if through the solution, the team is achieving the state of the art (SOTA). Although these two ways to frame the problem are considered, how much teams are able to determine the frame and then solve the problem by developing ML is constrained by limited resources such as time and budget. Trade-offs have to be made. Oftentimes this results in solutions that are "good enough" to meet set product metrics, but fall short of their potential.

The AI Ethics Brief    |    The MAIEI Learning Community Report

An alternative, and complementary approach is to start by framing the problem through the eyes of those who are experiencing it. Who is experiencing this problem? What is the problem? What is the need? What are the pain points? What are they hoping to achieve? A human centred design approach, which maps the different stakeholders and their needs, as well as desires and interests, helps keep these questions top of mind through the development of technology.

We recognized that there might be competing interests or perspectives when framing and trying to solve a problem. We recommend the facilitation of dialogue between the different actors and decision makers, and a collaborative process. This would help reduce the risk of working in silos and creating a solution that doesn't address the needs and desires of those whom the solution is meant to help.

When framing the problem, one must consider the consequences of solving the problem. Teams have to be conscious of the possible bias and self interest implicit in framing the problem. "Unless you truly care about those who are being negatively affected by the solution, you might actually cause a lot more damage than you intend." We looked at social media as an example: If a social media platform aims to engage users, and they succeed, what are the consequences? We should also be mindful that scaling up solutions can lead to unintended consequences. Christina offered a question she often encourages students learning ML to ask themselves: "What are the potential negative consequences of the ML? Even when I am trying to have a positive impact?"

Data

Data used in the development of artificial intelligence faces a myriad of problems during generation and usage that raise ethical issues. Approaching these issues from an interdisciplinary perspective could lead to solutions.

From the perspective of group members working in the tech industry, datasets used for machine learning are often auto-generated leading to silos between AI practitioners and those responsible for data generation. However, it was noted that collaborating across those boundaries is often simpler at smaller companies. Data can be auto-generated based on user clicks, for example, or by customer service representatives labeling data as they respond. Concerning human labelers, this leads to another often-discussed industry problem of how to scale; e.g., how to label the thousands of data points fast enough by hand. There is concern among industry and non-industry members alike that this leads to poor treatment of human labelers, such as the case documented at Facebook[1]. When industry members try to, for example, initiate conversations with people doing the data generation/labeling or data scientists responsible for cleaning data into a formal dataset, this adds additional steps in the product development process; tension forms when you are viewed as slowing down the process and this is often discouraged.

As our interdisciplinary group discussed these issues, a few common themes came up as a better way forward. We suggest that larger tech companies learn from smaller companies and
support collaborating across roles and stages in the AI development process. Our view is that having more conversations across roles and stages in the process will create empathy for those involved, implicitly or explicitly, in the data generation process. To illustrate this need, one of our group members presented a hypothetical situation within the criminal justice system: when an individual misses an important appointment with someone in the system for any reason (for example, because they did not have child care available, or they were ill), this counts against them and is recorded as a missed appointment in the system, generating data. This context is then lost later on when that data is taken at face value during AI product development.

Models

Once the training data is gathered, the next phase is model training and evaluation. The model development team, usually consisting of machine learning engineers and data scientists, proceeds to build and evaluate models. It is in this phase of the machine learning development life cycle where the efficacy of the solution is determined. Our interdisciplinary group is especially interested in discussing and understanding this phase of the process and the issues that arise.

The group believes that close conversation between the model development team and decision-making stakeholders would be greatly beneficial in determining what actions to take with the model outputs. Obtaining perspectives from experts outside of computer science and software development often leads to a more complete and nuanced understanding of the implications of making a decision based on a model's output. This, however, seldom happens in practice. In highly specialized teams, model evaluation is often performed only by the data scientists and machine learning engineers who tend to lean on established metrics (for example precision, recall and F scores), without considering downstream effects (for example, how a purchase recommendation changes the user's overall experience on the website). This siloed process then heavily relies on the quality assurance (QA) team to ensure consistency across discrete technical teams (for instance, data, front-end and back-end teams).

An additional aspect of model evaluation that the group also explored is model explainability. There is a general trend of machine learning models getting bigger in size (for instance, the recent rise of transformers-based large language models) [2]. There is a growing tension between opaque efficacy and explainability. Even when the model "performs well enough" as described in the previous paragraph, engineers often miss or ignore embedded biases. When engineers don't fully understand why the model performs well, they may fail to address issues when the model's performance degrades over time. The group also acknowledges that many machine learning teams are limited by access to resources, but still urges AI practitioners to plan early and make room for nuanced model evaluation in the development plan that (1) gathers a diverse set of perspectives for how the model affects end-users and (2) invests in explaining why the model performs well.

Testing & Feedback

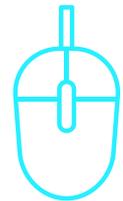

By the time a model reaches the testing and feedback point in the pipeline, its success is measured by how efficient it is: Testing generally looks to answer whether your model can do what it's "built" to do as quickly as it can. From the perspective of group members working in the tech industry, what could be an iterative, evaluative process is often stifled by deadlines, budget, and staffing resources. This leads to less human testing and more automation in testing. Once a project goes into maintenance mode, much more of the testing (or maintenance) becomes automated.

Given these limitations in resources, one might want fully automated testing, but at the moment technology hasn't advanced enough to allow organizations to do so responsibly. The question we raise is "Should we even want fully automated testing?"

If the ideal is to be perpetually feeding back and reevaluating your model and your data throughout the development process, then it is the group's opinion that AI is not exempt from the rigorous testing standards of other technologies. Considering the scale that AI could operate on, testing beyond efficacy and towards how well the ML tool helps users should be a high priority instead of seen as a nuisance. We suggest doing a study early on to get a benchmark and another study later on to see improvement of the aspects tested, like accuracy in prediction. As with any technology, there are multiple stakeholders. It's important to ask: "Who are we looking to benefit?" Incorporating elements of human-centred design at the testing and feedback stages of the pipeline allows for better understanding of the users and stakeholders involved in the development of the model. Asking: "Who is the end user?", "The person interfacing with the system?", "The communities impacted by the decisions made as a result of output of these systems?", and "Does the testing and feedback reflect these end users?" Throughout the journey, the users should have a say in whether the ML system is useful.

The involvement of users and multidisciplinary testing throughout the development process is important because it reduces the risk of producing something that is not beneficial. For example, a design that may seem good and useful to a limited team may fail when implemented in the real world because they didn't account for colour blind accessibility.

Even if you don't have a team that accounts for every single possible perspective, you could receive valuable input through multiple avenues of testing (engineer, user, designer, etc.). Having a multidisciplinary team as well as testing with diverse users in the loop will help finding balance between speed, efficacy, and other necessities such as accessibility and explainability.

Implementation

From the perspective of group members working in the tech industry, there's a lot of discussion about whether or not certain machine learning algorithms are abusive or coercive, but little debate on the accountability or restraints on how these are being implemented. It was noted that there are very few oversight bodies and most have limited powers, and most of the legislation at the moment focuses on data collection and privacy.

Our discussion reflected how keeping ML systems afloat requires an interconnected and interdisciplinary network and infrastructure (such as servers and hardware). Understanding how these pieces come together highlights how fragile the system really is. Monitoring and logging against failure once these systems have been implemented is important because it ensures the stability of the prediction and the outcome. It's important to account for harm not only when systems go wrong, but also when they go down. We often think about the negative implications of a poorly designed model, but we also need to consider that when the model does do good, failure of that system can now have drastic consequences for those who rely on it. It might not seem very important if you consider a movie recommendation engine going down for a little while, but system failure becomes critical in healthcare; a failure in a system that triages suicide risk might lead to missing potential cases.

We noticed that we came full circle in that we can't talk about implementation without answering the questions of problem framing: "Who and what is the ML being designed for?" Poor design is not just an aesthetic failing; bad design can have potentially fatal consequences (think of car systems and crash prevention), and manipulative designs can influence large scale consumer behaviour (like how people interact with and react to content on social media platforms), and can obscure users' understanding of, for example, how their personal data is being used.

Moreover, it's important to be mindful that implementing ML systems outside the context where they were developed might result in unintended consequences. For example, a face-recognition system trained to identify facial expressions in a North American population might not be as accurate when used with an Asian population. Also, consider that technology is not constrained by political boundaries, but policy is; what's protected in the EU might not be, for example, in South America. Finally, we can't always rely on the integrity of organizations developing these technologies.

### The way forward

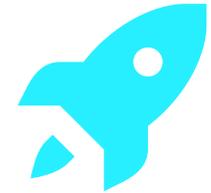

discipline mentorship, I've come to value those who bridge the gaps of communication between people with different perspectives. Mentorship is especially shown to be effective in increasing diverse participation in industries that have traditionally marginalized or excluded certain communities or knowledge domains, so I want to continue exploring ways of facilitating these kinds of personal, knowledge-sharing relationships.

### Tiziana

There is power in narrative and in sharing our diverse experiences. I will continue to encourage and facilitate conversations such as the one we had to produce this article. I am confident in saying we all learned from each other and have gained a new understanding of how technology, especially AI, benefits from a multidisciplinary approach and by considering: "What is the problem it is trying to solve?", "Is AI the appropriate tool to solve it?", and "Who is impacted in its development and deployment?" I will continue to document my learning journey as I explore the intersection of Human Centred Design, AI, and ethics.

### Brittany

Tech workers who care about these issues and are making the effort to inform themselves have a unique opportunity to directly raise awareness among their coworkers. By starting to discuss these issues—whether in study groups with coworkers, team meetings, one-on-one conversations, or hallway and lunch conversations—encouraging awareness and making space for fellow tech workers to start thinking about these issues is a crucial piece in addressing them.

### Mitchel

In the coming years, multidisciplinary development must be viewed as a preemptive risk-mitigation strategy. Business leaders and other supporting professionals, such as lawyers, need to help developers to understand the benefits that will be reaped in the long term from prudent planning and intentional and holistic development. Educating stakeholders on the salutary effects of multidisciplinary collaboration and creating a workable plan that takes into account the extra resources required for such a collaboration while balancing those extra demands with the realities of the development cycle will be a necessity. In any venture, there will be risk and it is the responsibility of these supporting professionals to help educate and mitigate these risks. It is my hope that as multidisciplinary collaboration becomes a more regular development consideration, the likelihood that it becomes integral to all decision making increases.

### Wilson

Tech workers who wish to improve the status quo might find themselves drawn to grassroots open-science initiatives, for example BigScience: the Summer of Language Models 21 [3]. These initiatives are multi-institution, multi-country, and entirely transparent to the public—which heralds unparalleled diversity of thought and perspectives.

### Christina

I was brought up in a culture that values specialization and highly technical expertise. Having completed an interdisciplinary degree and worked for an organization that promotes cross-discipline mentorship, I've come to value those who bridge the gaps of communication between people with different perspectives. Mentorship is especially shown to be effective in increasing diverse participation in industries that have traditionally marginalized or excluded certain communities or knowledge domains, so I want to continue exploring ways of facilitating these kinds of personal, knowledge-sharing relationships.

Tiziana

There is power in narrative and in sharing our diverse experiences. I will continue to encourage and facilitate conversations such as the one we had to produce this article. I am confident in saying we all learned from each other and have gained a new understanding of how technology, especially AI, benefits from a multidisciplinary approach and by considering: "What is the problem it is trying to solve?", "Is AI the appropriate tool to solve it?", and "Who is impacted in its development and deployment?" I will continue to document my learning journey as I explore the intersection of Human Centred Design, AI, and ethics.

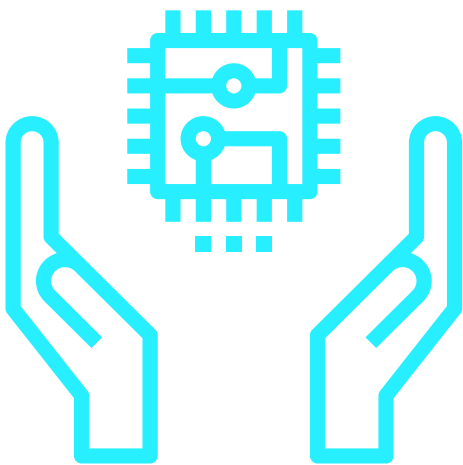

# 04

# FACEBOOK AND THE DIGITAL DIVIDE: PERSPECTIVES FROM MEXICO, MYANMAR AND INDIA

**Samuel Curtis, Shreyasha Paudel, Sofía Trejo, Tiziana Zevallos**

This chapter explores the methods by which Facebook has deployed its Free Basics program and other zero-rate data initiatives across Mexico, Myanmar, and India, in ways that have failed to meet the desires or needs of these respective communities. In Mexico, we observe that Facebook has usurped endeavors to expand meaningful connectivity by locally-based initiatives to roll out an application that does not meet the linguistic needs of the country. In Myanmar, we witnessed Facebook's negligence in providing the resources necessary to moderate the user-published content on their platform, allowing hate speech to fester. In India, we discovered how, after having early efforts to roll out Free Basics stifled on the basis of violating net neutrality, Facebook effectively exploited regulatory loopholes and close relationships with Prime Minister Modi to advance their infrastructure agenda. Time and time again, Facebook's one-size-fits-all approach to expand access to its platform (for advertising revenues) has led to detrimental outcomes for netizens.

Introduction

What began as a social networking experiment in 2004 to connect university students using "face book" directories has transformed into one of the (if not *the*) most influential companies to ever have existed.

Today, not only has Facebook's social networking platform captured a user base of over one third of the world's population, but in many parts of the world, the company has become synonymous with digital connectivity and the internet itself. Surveys conducted amongst mobile phone users in Nigeria and Indonesia, for example, demonstrated that respondents were more likely than not to agree with the statement: "Facebook is the internet."[1].

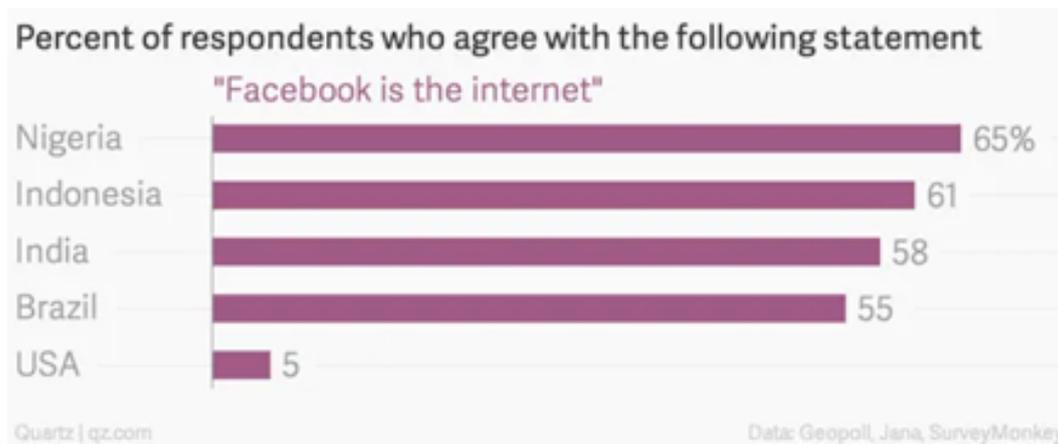

Facebook's stated mission is to "give people the power to build community and bring the world closer together." [2] In recent years, this has been highlighted by Facebook's foray into emerging technologies, such as the acquisition of augmented reality hardware company Oculus, and the development of blockchain-based payment system Diem. As captivating as these ventures may be, Facebook has in fact achieved more—in terms of substantive progress towards its mission—through its sweeping effort to provide internet services to countries within Asia, Africa, and Latin America.

Coinciding with the publication of Zuckerberg's 2013 white paper "Is Connectivity A Human Right?"—in which he argues the affirmative case—Facebook launched Internet.org, a partnership between Facebook and six telecommunications hardware and service companies, aimed at providing affordable internet access "to connect the next 5 billion people." [3]. Functionally, Internet.org became a vehicle for a partnership with telecommunications infrastructure and service providers to create low- or no-cost mobile data plans, conditional on users accessing only a limited number of services through the application available at Internet.org. The application would permit free access to Facebook.com, in addition to a number of other (often US-based) sites and/or services, with accessibility depending on the SIM operator used and the region in which it was accessed [4]. In 2015, Facebook began receiving criticism that Internet.org's content restrictions were violating net neutrality, facing prominent backlash in India. Likely in direct response to this backlash and to the gravitas the name carried, Internet.org was renamed in 2015 to Free Basics, with Zuckerberg remarking, "we want to make it clear that the apps you can use through Internet.org are free, basic services."

In July 2017, journalists from the nonprofit Global Voices published a report that investigated how Free Basics was meeting the needs of its users in six different countries: Colombia, Ghana, Kenya, Mexico, Pakistan and Philippines. These journalists concluded that the service "does not meet the linguistic needs of target users … features little local content, but plenty of corporate services from the US and UK … doesn't connect you to the global internet — but it does collect your data … violates net neutrality principles … [and] will not meet the most pressing needs of those who are not online." [5].

Relatively unscathed by critique, the Free Basics program continued to grow in geography and usership, reaching 65 countries as of July 2019 [6]. In spite of its expansive geographic reach, Facebook offers close to no information on the program—its website lacks any information pertaining to its terms, offerings, and limitations, nor describes how these differ from country to country [7]—and only a handful of studies (in a limited number of countries) have performed such investigations.

In countries where Facebook operates as the *de facto* internet forum, it is impossible to disentangle connectivity with the influence of the social media network's recommendation and content-moderation algorithms. Because these algorithms demand human oversight and consistent tuning, negative effects may be compounded in countries towards which Facebook dedicates fewer resources towards fine-tuning the performance of these algorithms—adapting them to localized language, politics, norms, and mores—these algorithms perform comparatively worse [8]. And while Facebook has taken some action to enhance self-governance through the establishment of its Oversight Board, the Board's purview only pertains to content moderation within the social media platform [9]. The Oversight Board's mandate does not cover the organization's wider strategic pursuits pertaining to infrastructure and connectivity, which funnel internet users onto the platform and subject them to algorithmically-moderated content in the first place.

Throughout this investigation, we must keep in mind Facebook's most valuable asset: its users' data. Wherever Facebook is accessible, the company profits from data extractivism — collecting and profiting off of their users' locations, demographics, interests, behaviors, and connections [10]. This is evidenced by the lion's share of Facebook's revenue: advertisements [11]. In 2020, Facebook closed its fiscal year with $85.9 billion in revenue and a market value of $800 billion, with 97.9% of its global revenues coming from advertising [12][13].

Though Facebook has often cast the Free Basics program as a purely humanitarian ambition, users' data readily becomes a product for retailers, services providers, and politicians willing to pay for targeted advertising. User data is also an invaluable resource for the development of AI systems. Earlier this year, in fact, Facebook announced a project titled Learning from Videos, designed to "automatically [learn] audio, textual, and visual representations from the data in publicly available videos uploaded to Facebook." [14]. When asked by The Verge journalists whether users had to consent to having their videos used for this research or if they could opt out, the company referred to its Data Policy, which states that users' content can be used for "product research and development." [15].

The sections that follow explore how Facebook's deployment of Free Basics and other zero-rate data initiatives has evolved in three different countries—Mexico, Myanmar, and India—and how Facebook's escapade to provide (limited) internet to the masses has fallen short of their goal of empowering communities and bringing the world closer together. On the contrary, the approach to providing moderated public fora falls short of meeting the linguistic needs of target audiences, and has perhaps even played a role in exacerbating social and political tension rather than serving as a space for mediation.

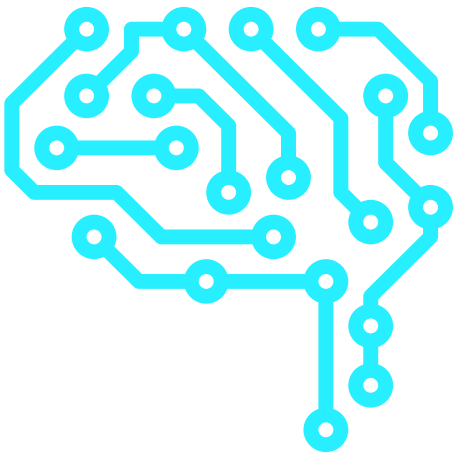

# MEXICO

In this section, we will explore Facebook's expansion in Mexico and links between the company, Mexico's Government, and local mobile network operators. We will begin with a brief overview of Mexico's context, in relation to cultural diversity and connectivity. We believe this information will help readers better understand the analysis we present of the Free Basics programme and other Facebook-related initiatives in the country. We will close this section by highlighting local telecommunication projects which align with the visions, interests, and aspirations of local communities, which present alternative paths for the design, ownership, and development of technology.

Mexico is a multicultural country composed of several distinct nations, many of which self-identify as *pueblos originarios* (first people), a term commonly preferred by them over *pueblos indigenas* (indigenous people). At present, 68 indigenous languages are spoken in the country, the most prominent being *Nahuatl,* which is spoken by nearly a quarter of indigenous language speakers [16]. It is estimated that almost 25.7 million Mexicans (21.5% of the population) self-identifies as indigenous and that 7.2 million speak an indigenous language [17]. In particular, four states—Oaxaca, Yucatán, Campeche and Quintana Roo—are composed of a high concentration (around 50%) of indigenous population [18]. Of indigenous-language speakers, more than 60% live in rural areas, mostly in localities with less than 2,500 inhabitants, and one out of ten do not speak Spanish (they only speak one or more indigenous languages) [19]. Indigenous-language speakers' literacy levels [20]are much lower than the national average (on average 5.7 years vs 9.1 years in school), and this gap is larger for female indegenous language speakers [21].

In terms of connectivity, universal internet access became a constitutional right in Mexico in 2013 [22]. In the time since, the Mexican Government has designed and implemented diverse strategies to guarantee this right and reduce the digital divide, and from 2015 to 2019, the number of internet users in the country rose from 62.4 million to 80.6 million, reaching 70.1% of the population older than 6 years of age [23].

The levels of connectivity, devices used to go online, and uses of these technologies vary throughout Mexico and are linked to social, economic, and geographical factors. For example, the vast majority of users in the country access the internet via mobile phones (93.5%), while only about one third has access to laptops (33.2%) or desktops (28.9%) [24].

Furthermore, of users who go online using mobile phones, more than 90% do so using data plans, instead of Wi-Fi. Despite the increase in access to the internet, the quality of connectivity in the country remains poor. More than 50% of users (40.4 million) report having slow connection, while 38.6% (31.1 million) experience frequent connectivity loss [25]. In addition to these issues, the country is still short of closing the digital divide. Figures from the National Institute of Statistics and Geography (INEGI) show that while connectivity has risen in the country, rural areas have remained about 30% less connected than urban ones [26].

The process of expanding connectivity and closing the digital divide in Mexico has involved conversations with two of the country's Presidents with Facebook's founder, Mark Zuckerberg. The first of these conversations happened in September 2014, when Zuckerberg met with President Enrique Peña Nieto (in office 2012–2018) to discuss the introduction of Internet.org in the country [27]. Only a year later, in December 2015, Facebook's Free Basics program launched in Mexico. Initially, the only provider of the service was the network operator Virgin Mobile. Then, in 2017, Telcel (the main telecom operator in the country) was added to the program [28]. The benefits and impacts of the Free Basics program in the country have not been widely studied. However, a report produced by Global Voices [29] analyzed the Free Basics program (May–June, 2017) in six countries in Africa, Asia, and Latin America (including Mexico) to evaluate its alignment with the program's goal of narrowing the digital divide by connecting disconnected people around the world. The results of Global Voices investigations in Mexico revealed that access to the Free Basics program between 2015 and 2017 was limited for two main reasons. First, the lack of vendors of Virgin Mobile SIM cards in rural areas and in low-income localities. For example, there were no Virgin Mobile shops in Chiapas and Oaxaca, two of the states with less connectivity in the country. The second reason was the lack of compatibility between Virgin Mobile network and diverse smartphone models [30].

Regarding the services provided, Facebook's Free Basics app in Mexico offered access to 16 default sites/services (tier-one), which included Facebook and Facebook Messenger. Of the tier-one sites available, via Virgin Mobile, only two were hosted in Mexico: OCC Mundial, a job recruitment site; and Crea Comunidades de emprendedoras, an NGO for women entrepreneurs. Only one site (tier-one) was offered via Telcel: Fundación Carlos Slim, a foundation of Carlos Slim, who is the CEO of Telcel. Furthermore, eleven of the tier-one sites were owned by private companies, most of them based in the US, such as ChangeCorp (three sites), Johnson & Johnson, and ESPN. Only one Mexican news outlet, the newspaper El Universal, was included, while access to numerous international news outlets was possible: Deutsche Welle (in Spanish), Voice of America (in Spanish), BuzzFeed News, Various editions of Xataka (Latin American technology site), Reuters (in English), Necochea news (news outlet in Argentina), TrendyRammy (Nigerian entertainment news site) and 24/7 News Nigeria (Nigerian news site). No social media sites, besides Facebook, or email services were accessible.

Global Voices investigation reported that, despite Mexico having 68 indigenous languages, Free Basics was available in only Spanish and English and required some English skills to operate (as several site names and contents were in English). In addition, their investigations showed that the app did not seem suitable for people with diverse levels of literacy, as the content was text-heavy and in many cases sites removed images and videos to throttle data usage. Furthermore, the app was not well adapted for disabilities; for example, it did not offer voice and audio assistance.

In relation to the target audience, it was reported that Virgin's promotion of the service seemed to focus on millennials and young people who frequently used Facebook's platform and wanted to avoid exceeding their data limits. The Global Voices report concluded that the Free Basics program: (1) did not meet the linguistic needs of target users; (2) featured an unbalance of key sites and limited the addition of small independent platforms to the service by setting unique technical requirements (difficult to meet with limited resources); and (3) was giving Facebook access to data of non-Facebook users, since all the traffic of the Free Basics app passes through special Facebook-owned-and-operated proxy servers.

In addition to the introduction of Free Basics, Facebook has been included in a wide range of zero-rate data initiatives in Mexico for many years. For example, in 2015 Facebook launched Facebook's Switch option, which enabled customers of AT&T, Nextel, Iusacell and UNEFON to navigate Facebook using a text-only mode at no additional charge [31]. Currently, most mobile operators in Mexico, including Telcel, UNEFON, and Movistar, offer "unlimited" access to many of Facebook's owned platforms including Facebook, Facebook Messenger, Instagram and WhatsApp. Digital Rights organizations, such as Red en Defensa de los Derechos Digitales and Derechos Digitales, have been outspoken in their criticism of zero-rate initiatives, as they provide undue strategic advantage to big technology companies [32][33] Besides these initiatives, there has been fraud in the form of online profiles and sites offering "unlimited" Facebook access in exchange for bank transfers. The problem was so serious that in July 2018 the Mexican Government (PROFECO) issued a press release to warn consumers of "Free Facebook" scams in the country [34].

With the transition of power from Enrique Peña Nieto to Andrés Manuel López Obrador, in December 2018, the telecommunications landscape in Mexico shifted, but the inclusion of Facebook in national conversations around the digital divide remained. For example, on June 18th, 2019, President Andrés Manuel López Obrador tweeted a video of a virtual conversation he had with Mark Zuckerberg. In the video, the President invites Facebook to collaborate with the Mexican Government to reach the goal of providing internet access to all Mexican citizens, particularly to marginalized communities, by creating a non-profit partnership.

The video sparked widespread criticism, since the country has many existing projects and organizations which, for years, have been working in collaboration with local communities and indigenous groups to create autonomous and communally-owned telecommunications services. It is important to highlight that the Mexican Constitution (Article 2) [35] establishes that the Mexican Authorities (at federal and local level) have the obligation to extend communication networks to enable the integration of indigenous communities; as well as the obligation to establish conditions for these communities to own, operate and maintain such networks. Furthermore, public policies which relate to indigenous communities should (by law) be designed and operated in conjunction with them.

The outcome of Obrador and Zuckerberg's conversation remains unclear. The current National Programme for providing internet access to all Mexicans is called Internet para Todos (Internet for Everyone) and it's run by an autonomous government body with the same name [36]. Since the 28th of May 2020, ACTLÁN Redes, a mexican telecom company, holds the Public Private Partnership (PPP) contract for the design, financing, deployment, operation and commercialization of the federal mobile broadband network, called Red Compartida [37]. The government expects ACTLÁN to connect 92.2% of the country by 2024 [38]. It is important to highlight that even though the Government has set clear goals for expanding connectivity in the country, it lacks a National Digital Strategy to guide the process. Since 2019, announcements have called for the development of such a Strategy [39]; however at this date it is unclear when it will be presented or implemented.

Other paths possible

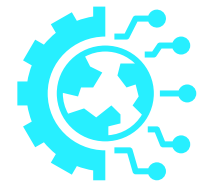

Treating Telecommunications services as commercial rather than social enterprises implies that connecting remote communities with small populations is not profitable, hence not viable. For example, GASMA (an organization representing the interests of mobile operators worldwide) estimates that for localities to be commercially viable they should have at least 3,000 active subscribers monthly. Hence, they should have at least 5,000 inhabitants and should be in a location where other viable communities exist within a range of 25km [40]. These economical restrictions limit the human right to communication [41] and reduce the community's abilities to exercise this right in ways which align and respect their values and principles. These issues are particularly relevant, as indigenous communities in Mexico have the right (Article 2 of the Mexican Constitution) [42] to autonomously determine their social, political, and economic structures; and the right to live and communicate according to their own practices, procedures, and cultural identities.

To counterbalance the limitations to communication posted by commercial enterprises, several organizations in Mexico work to create communitarian radios, mobile services, internet services, and intranets, which are run and owned by local communities. For example, Telecomunicaciones Indígenas Comunitarias (Communitarian Indigenous Telecommunications Civic Association, TIC A.C.) [43] is a civil association integrated by indigenous and rural communities, along with an operations team which accompanies people and communities in the process of constructing, maintaining, and operating their own telecommunication networks. In July 2016, TIC was granted the first ever Social Indigenous Concession to manage and operate autonomous telecommunication; their Concession includes five states: Oaxaca, Guerrero, Puebla, Chiapas and Veracruz.

Another organization, which works alongside TIC, is Rizomatica [44], which supports communities to build and maintain self-governed and owned telecommunication infrastructures [45]. Rizomatica also engages with policy and regulatory advocacy, develops free and open-source software, and produces research in collaboration with the Centro de Investigación en Tecnologías y Saberes Comunitarios (Center for Research in Technology and Community Knowledge, CITSAC).

All of these projects and organizations aim to accompany communities in realizing their aspirations in relation to communication. They support the development of local contents; they design communications strategies to help communities defend their territories (material and immaterial); and they work on policy and regulation to support communities in exercising their right to communication. These projects are only a few examples of a vast network of community networks whose connections, ideals and practices expand far beyond Mexico, into Latin America and the rest of the world [46].

Most importantly, by aligning with the principles of autonomy, self-determination and solidarity, projects like those above are a counterpoint to the capitalist, extractivist, and exploitational practices and narratives which permeate the hegemonic discourse around technological development. They are clear examples that other ways of developing technology for social connectivity and cohesion are possible. They present us with the opportunity of communal ownership, management, and control; with knowledge of how to develop projects led by local communities; and with tools to attend to local needs and aspirations in ways which respect and help nurture diverse ways of life.

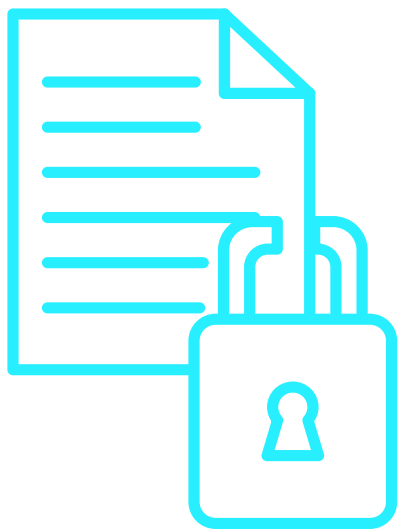

# MYANMAR

In this section, we will explore the expansion of Facebook in relation to the political and social context of Myanmar. We evaluate how the government's initiatives to increase connectivity and ease media censorship provided fertile ground for Facebook's rise, leading to Facebook becoming synonymous with the Internet. As Facebook became the main source of news, we discuss how Facebook acted as medicine for some, in providing people with a space to easily access and share different viewpoints and document events; and how Facebook acted as poison for others, as it worked under the assumption of a level of digital literacy which allows people to fact-check information. We also observe how Facebook's algorithm likely amplified selective exposure to information and engagement with dangerous speech, and its role in heightening polarization.

Further, we explore evidence which suggests Facebook acted negligently in prioritizing the expansion of its network without considering and developing the appropriate measures to prevent and mitigate the unintended effects of doing so. In this section, we place value in research studies and reports that consult the perspectives and lived experiences of people in Myanmar.

Following the election of a nominally civilian government in 2010, after decades of military rule, the landscape of internet freedom and access in Myanmar began to change [47]. In 2012, an estimated 500,000 people, less than 1% of the population in Myanmar, had Internet access [48]. In an effort to reach 30 million phone lines, in a country with a population of 51.41 million, by 2016 (approximately 58 percent of the Myanmar population) and connect four million new internet users within the first year [49], the government of Myanmar lifted the monopoly of the Myanmar Post and Telecom (MPT) and issued concessions to international providers to build networks across the country [50]. As a result, SIM card prices dropped in half, from 500,000 kyats (US$625) to about 250,000 kyats (US$312) [51]. By 2014, SIM cards could be bought for less than US$1. A year later, Myanmar ranked as the world's fourth-

astest growing telecommunication market [52].

During this time, social media tools gained prominence. Among five platforms (Twitter, Friendfinder, Netlog, Google+, and Facebook), Facebook became the most popular, with an estimated 80% of the country's internet users having a Facebook account [53]. To increase its share of this growing market, Facebook introduced its Free Basics program in 2017 and provided a Burmese language version of its platform [54]. Facebook soon became synonymous with the Internet and a primary source of news as 18 million people in Myanmar used Facebook by 2018 [55].

Facebook's rise coincided with easing government pre-publication censorship on local media and eroding trust in foreign media. For a country with a history of "ethnic conflict, colonialism, military rule, and media censorship," [56] Facebook and the Internet were mostly unregulated. Facebook became a medium to circumvent government's indirect control of domestic media and a space where diverse information became available [57].

To understand Facebook's expansion in Myanmar, it is important to understand the country's historical and ethnic context. Myanmar, also known as Burma, was under military rule from 1962 to 2011. During this period, the government imposed strict state media censorship, and has been accused of suppressing dissent as well as gross human rights abuses [58]. A gradual liberalisation beginning in 2010 led to free elections in 2015 where Aung San Suu Kyi, a veteran opposition leader, was elected and then named State Counsellor. Since August 2017, Myanmar's army known as the Tatmadaw has lead operations against the Rohingya, a stateless Muslim minority in Rakhine State, leading to 1.3 million Rohingya fleeing to neighbouring Bangladesh. This raised the Rohingya's plight of over five decades into international prominence. Aung San Suu Kyi faced international pressure to intervene in the Rohingya crisis to no effect, signaling the continuing power hold of the military in Myanmar [59]. War crimes and crimes against humanity against other ethnic minorities (there are 135 official ethnic groups in Myanmar) [60] have also been carried out by the Tatmadaw in this timeframe [61]. In February 2021, the Myanmar military staged a coup nullifying the results of the November 2020 elections and transferred power to the commanding officer of the Tatmadaw, Min Aung Hlaing [62].

Facebook became a space where the proliferation of extreme speech and disinformation against ethnic minorities, especially the Rohingya, were made widely available, easily accessible, and difficult to contest [63]. Although Facebook's Community Standards state that hate speech is not allowed, "much of the anti-Muslim rhetoric posted on Facebook in Myanmar from 2012 onward" clearly met the criteria to be removed, but remained on Facebook [64]. Notably, in 2017, the Tatmadaw launched a military operation accompanied by a social media campaign on Facebook against the Rohingya [65]. The operation resulted in the death of thousands of Rohingya and the displacement of 700,000 Rohingya [66]. At the time, United Nations officials called it "a textbook example of ethnic cleansing." [67].

"Internal conflicts have simmered for decades in Myanmar. More than a dozen ethnic armed groups are fighting for autonomy against Myanmar's national army, known as the Tatmadaw." [68]. Although multiple armed groups have operated in Myanmar since 1948, and utilized different communication strategies, social media has significantly lowered the barriers and increased the speed of communication. It has allowed conflict actors to frequently shift media strategy and tactics, as well as provided a powerful tool to pose as legitimate states, by mimicking the symbols and structures of recognised states on their online profiles. "The desire to pose as states is not new. [...] What is new is that the messages can be conveyed immediately and at little cost to huge audiences." [69].

A research study [70] published in 2020, aimed to understand how people in Myanmar use Facebook through a series of twenty-two in-depth individual interviews and four focus group discussions. These ranged from five to twelve participants each, with experts, average Facebook users, and nonusers in Myanmar [71]. The study found that "people in Myanmar use Facebook for a variety of reasons, that they look at international and domestic sources, and encounter many different kinds of information—including extreme speech—on the platform. Most participants also [said] that they prefer sources that confirm their views." [72]. In response to whether Myanmar is better off with or without Facebook, a participant captured the essence of the research findings sharing a Burmese saying: "[Facebook] will be medicine if you know how to use it, but it will be poison if you don't know how to use it (Thone tat yin say, mathone tat yin bay is a Burmese saying used here in reference to Facebook)." [73]. Using this analogy, we discuss how Facebook acted as medicine and how it acted as poison in Myanmar.

Facebook acted as medicine by allowing different viewpoints to be shared, giving a space to those who couldn't share their stories through traditional media by circumventing censorship and providing the documentation of human rights violations. The research shows that although the reasons why people used Facebook varied, most agreed it was an easy source of information. Most claimed to use Facebook for political news, and for the majority of respondents, it was their primary source of news, especially for those in rural areas. Facebook users recognized that the platform can be used to spread disinformation and extreme speech, but that some might not be identifiable as such because they do not explicitly use hate words.

The methods users employ to determine the veracity of information (or fact-checking) span from using their instinct or intuition, or analyzing the tone of the message to determine if it's "fake" information, to double-checking viral posts by corroborating with other sources. The research also suggested that Facebook users in Myanmar are aware of their preference towards sources that confirm their views. However, the majority of participants in the study would still prefer privately owned local media that might confirm their perspective, and were dismissive of accounts by international media as they might challenge them [74]. Participants reported using Facebook to circumvent censorship and "official" narratives that fail to represent the full extent of events in Myanmar, and to access diverse perspectives "from real people." [75]. Some argued that Facebook gives voice to people of minority ethnic areas and offers information other outlets do not cover. This suggests that Facebook allowed for the reporting of the conflict and violence directed at the Rohingya [76] that otherwise might have gone undocumented [77]. Reflecting on Facebook's role of dividing or uniting people, users said the platform mirrored an already divided society, "bringing the country's existing ethnic cleavages into full view." Both online and offline, hostility towards the Rohingya is communicated the same [78].

Facebook acted as poison as it worked under the assumption of a level of digital literacy and left people to their own devices to fact-check information, and its algorithm design has resulted in selective exposure to information and engagement with dangerous speech. For those not on the content creation but content consumption end, the platform places the responsibility (for the most part) on users to verify, flag, or counter disinformation, fake accounts, extreme speech, and disturbing images. Facebook allows users to create an online space with content they agree with, and avoid information that would challenge their views, resulting in selective exposure to information, also known as echo chambers [79]. Even when individuals seek diverse perspectives, observations have pointed out that the algorithm prioritizes the user's implicit preferences by hiding information that may challenge the user's views [80]. This content prioritization is most likely due to the algorithm trying to maximize the time users spend on the platform [81].

Research suggests that such online echo chambers might significantly contribute to heightening polarization and diminishing democratic deliberation [82]. Therefore, if individuals engage with one-sided information, which can include fabricated information or extreme speech targeting a particular group, it can lead to more anger and increased ethnic violence [83]. In the context of Myanmar, it is important to highlight the cultural and usage norms of the platform. Another exploratory study conducted in Myanmar [84] found that "in order to be kept informed, the majority of [study participants] employed the same strategy to gather information: Friending. While Facebook states that "You should send friend requests to friends, family and other people you know and trust on Facebook" (Facebook 2017), many Myanmarese [study participants] talked of adding strangers as Friends to gather information to, in their understanding of the Facebook, funnel into their News Feed." [85]. According to Hootsuite, "the Facebook algorithm evaluates every post, scores it, and then arranges it in descending order of interest for each individual user," taking into consideration the relationship (person, business, news source, or public figure), content type (video, photo, or link), popularity (how people, especially your friends, are reacting to the post), and recency (newer posts are placed higher) [86].

Therefore, in Myanmar, it is likely that users are seeing content posted by Friends first, whether they know them or not. The question we raise is: "What is Facebook's role in preventing it from becoming poison?" and if its prevention efforts fail, "What is Facebook's role in producing and administering the antidote?" The evidence so far suggests Facebook acted negligently in prioritizing the expansion of its network without considering and developing the appropriate measures to prevent and mitigate the unintended effects of doing so [87].

Facebook's quick roll out of a Burmese version of its platform was key to its success. However, until 2018 Facebook had few Burmese-speaking staff. Concerned civil-society groups had been sounding the alarm of the company's unequipped approach to monitoring mechanisms, policies, and actions for years. However, it was only after the United Nations and international media criticized Facebook for its role in the displacement of Rohingya [88], and for its slow reaction to viral messages that incentivized Buddhists to arm themselves in preparation for Muslim attacks (and vice versa), that Facebook increased its number of Burmese language content monitors from 1 (in 2013) to 100 [89]. Nevertheless, the company has still been heavily relying on a technological monitoring approach that uses artificial intelligence to search for keywords for moderating content [90].

Since 2017, Facebook has responded to criticism by enhancing its capacity to remove content and remove accounts that violate its Community Standards [91], such as those by armed groups. Notably, in 2018 "after a UN fact finding mission had accused it of crimes against humanity with possible genocidal intent," the company cancelled accounts controlled by the Tatmadaw (Myanmar's armed forces) as well as banning its commander-in-chief Min Aung Hlaing, now Myanmar's de facto leader." [92].

Facebook states that between October and December 2020, it took action on 350,000 pieces of content containing hate speech in Myanmar, of which 99% were detected and removed before anyone reported it [93]. However, a report by the human rights group Global Witness states that Facebook has been recommending users to like pages that share pro-military propaganda that violates the platform's rules [94]. The investigation conducted in March 2021, found that by searching "Tatmadaw," the first five recommended pages were followed by almost 90,000 Facebook users and posted content that violated Facebook community standards [95].

An Update on the Situation in Myanmar, posted by Facebook on it's Newsroom site in April 2021, detailed the implementation of a specific policy for Myanmar under their coordinating harm and publicizing crime policy to "remove praise, support and advocacy of violence by Myanmar security forces and protestors from our platform." [96]. Facebook lists the following examples of content they will remove under this new policy: "A post with an image showing shots fired into a crowd with the caption:'"I support this!;' Posts explaining or arguing that lethal violence against people is right; Posts praising attacks that have been committed; An image of a person with a caption saying: 'I support the shooting of people.'" [97].

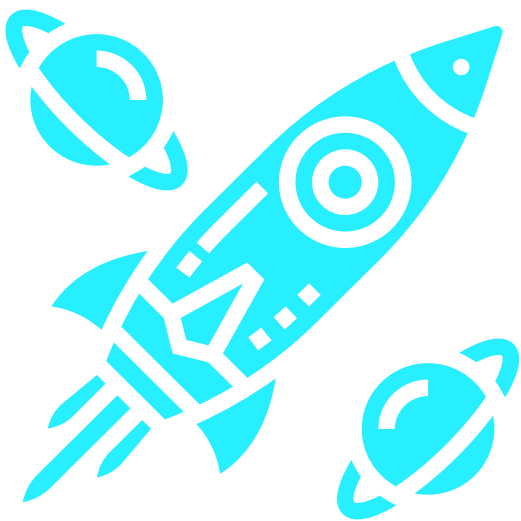

# INDIA

India is one of the largest markets for Facebook and a region prioritized by Facebook for strategic growth [98]. This section will look at initiatives led by Facebook in India to gain market and infrastructural dominance. In particular, we will look at the launch and subsequent failure of Free Basics in India and the more recent investment announced by Facebook for a payment platform. The case study situates these two events within the political and economic context in India and explores challenges in keeping large and powerful multinational corporations like Facebook accountable when they can easily leverage these contexts and adapt business strategy to advance their agenda.

Free Basics was launched in India as a partnership between Facebook and Reliance Communications, a major telecom company of India, in February 2015. The service was pitched as an initiative to improve internet access among poorer and rural Indians and was launched with free services for 38 websites [99]. The service attracted widespread criticism from net-neutrality advocates in India despite Facebook's efforts to market this as a humanitarian effort to provide internet [100]. The protesters, consisting mostly of engineers, entrepreneurs, and policy advocates organized themselves into a volunteer group dubbed Save the Internet (STI) and argued against Free Basics on the basis of net neutrality [101]. They claimed that Free Basics threatened "permissionless innovation" on the web, because it advantaged the Facebook-selected apps over any competitors [102]. Facebook defended its initiative, countering that "net neutrality considerations should not lead to a clamp-down on solutions for expanding connectivity to the most marginalised" [103]. Despite a large advertisement campaign by Facebook, STI advocates were able

o use blogs, comedy videos, and social media hashtags to mobilize the public to their cause, soliciting over a million emails and comments against Free Basics in less than a month [104][105] . Facing public outcry, several of Facebook's partners pulled out, and finally, in February 2016, Free Basics and other services like it were banned by the telecom authority of India on the basis of net neutrality [106].

This was a notable failure for Facebook. However, this David beats Goliath story can be best understood by examining the dynamics between Facebook, the STI advocates, and greater technopolitics within India. In their protest, STI advocates "emphasized their role and potential as technology creators and leveraged their privileged place in a contemporary India that imagines itself both high tech and global" [107]. The advocates' arguments swayed the public and resonated with the Digital India initiative pushed by the ruling party Bharatiya Janata Party (BJP) at the time.

Learning from its failure, Facebook has attempted to build a more amicable relationship with the government of India. Since 2016, Facebook has hired multiple people associated with the ruling party BJP in key decision-making positions [108][109] . One such example is Shivnath Thukral, a public policy director at Facebook who was a volunteer in Prime Minister Modi's reelection campaign in 2019. Former Facebook employees have disclosed that Thukral's key responsibility has been to ensure that an advocacy campaign similar to net neutrality does not happen again [110][111] . There have been multiple documented cases of Facebook removing posts critical of the government, while permitting posts by members of BJP even when the former abided by community standards and the latter violated them [112][113][114].

On the other hand, the government of India has been increasingly aggressive in regulating media, including internet companies. Studies have shown that censorship is on the rise in India [115][116]. India has also been falling in Democracy Index's global ranking: "India's global ranking slipped from 27th (in 2014) to 53rd (in 2020) as a result of democratic backsliding" [117]. In February 2020, the government introduced far-reaching regulations for social media content. These rules require social media companies, such as Facebook, to promptly remove content upon receipt of government orders, to assist law enforcement on cyber security-related investment, and to reveal the originator of any message or post [118]. Platforms' reactions to the Indian government's regulation and censorship have been mostly cautious [119] while critics, such as Internet Freedom Foundation, have responded with accusations of political control and censorship [120].

In this context, Facebook has announced another large investment to gain platform dominance in India. In April 2020, Facebook announced a partnership with Reliance Jio, India's top telecom carrier, with an investment of USD 5.7 billion [121][122][123]. The partnership will be focused on building payment infrastructure and working with local retailers and small businesses to bring them online [124]. This investment brings our story full-circle: Reliance Jio was launched in 2015 and was able to disrupt the mobile market in India and bring cellular data to billions of customers who did not have it previously—as was the goal of Facebook's Free Basics program. Jio did so by launching with free voice calls and cheap data plans, effectively forcing its competitors to do so as well. It also tied itself closely to the government's Digital India campaign and appealing to nationalist feelings in its marketing campaigns [125].

Within a year of its launch, Jio was able to acquire 130 million users, reduced data prices by more than 90%, and increased mobile data consumption in India by 600% [126]. Jio was able to offer the free data services because of the key infrastructure and spectrum that it had been able to acquire [127]. Competitors and critics allege that Reliance Jio was able to do so by taking advantage of regulatory loopholes and exploiting a close relationship with the Indian Prime Minister Narendra Modi [128][129]. The government has also repeatedly allowed Jio to renew and extend its promotional offers well beyond the stipulated deadlines even though other cellular operators have complained of predatory pricing and nepotism [130].

Internet advocates in India have criticized this deal and have accused both Facebook and Reliance Jio of being the gatekeepers of the country's internet infrastructure. They have also warned of potential data misuse, citing Facebook's previous scandals such as the Cambridge Analytica case. However, Facebook's second attempt at gaining platform prominence in India has been on much better footing and more aligned with the technopolitics in India. Criticism did not gain as much muster and the regulatory body approved the investment deal on June 23, 2021 [131].

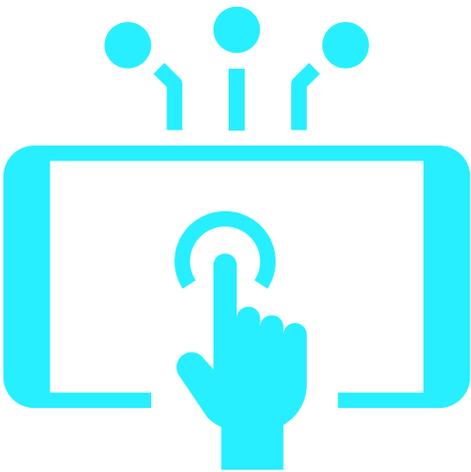

# CONCLUSION

This chapter has explored the methods by which Facebook has deployed its Free Basics program and other zero-rate data initiatives across Myanmar, Mexico, and India, in ways that have failed to meet the desires or needs of these respective communities.

Responding to accusations of Internet.org's violation of net neutrality in April 2015, Zuckerberg wrote: "To give more people access to the internet, it is useful to offer some service for free. If someone can't afford to pay for connectivity, it is always better to have some access than none at all." It seems that here, Zuckerberg himself confused Facebook for the internet. Rather than liberating the digitally unconnected with unfettered access to the internet, Facebook's zero-rate data initiatives primarily serve to build new streams for ad revenue by funneling digital (illiterate) newcomers into ill-resourced, algorithmically-curated, vitriol-prone social media spheres.

As individuals, companies, organizations and governments recognize connectivity as a right, these case studies stress the importance of ensuring that social media, and the AI applied within these spheres, meet communities' linguistic and social needs. This must be an ongoing, delicate, and deliberative process, not an afterthought. Based on the harrowing observations of only three of the over 65 countries in which Facebook implemented Free Basics, we advise against such cookie-cutter approaches towards bridging the digital divide in lieu of locally-based initiatives that are better suited to be aware of and reactive to the social, political, and economic characteristics of the communities in which they are based.

If the events of 2020 have taught us anything, it is that connectivity is directly linked to the protection and advancement of other rights such as access to information, education, work, and equal opportunities [132]. Cognizant that in many of these countries, data-extractive zero-rate data initiatives will continue to proliferate, we urge readers to reflect on a holistic approach to digital connectivity *as a right*—as it is widely becoming regarded in the 21st century—and what it should look like: How ought it to be guaranteed? Whose interests would be prioritized, via what mechanisms? Who should design and implement digital infrastructures and data governance strategies? How are these decisions being made now, and who will have a say in the future?

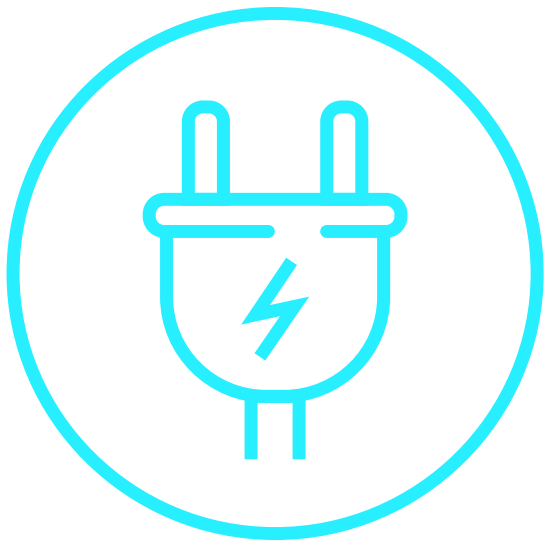

# FUTURE OF WORK

**Nanditha Narayanamoorthy**

Introduction

    One of the major transformations in the realm of human labour in the past decade has been the "emergence of digital labour platforms that have disrupted traditional models of business" and labour [1]. In the wake of automation and digitalization, "new" forms of employment are emerging [2], and the recent development of digital technologies has given birth to digital labour in the form of platform-mediated work such as "crowd work," "gig work," and other forms of on-demand labour with various stakeholders that provide opportunities for work [3]. Platform work has profoundly transformed the world of labour, by connecting workers with opportunities transcending geographical boundaries [4]. The platform economy, currently estimated at more than $4 billion [5], has now penetrated many parts of the Global North and South, and spread to numerous sectors and occupations including positions in digital services such as data entry, translation, design services, or location-based service in transport and delivery.

 Platform work is now characterized by the algorithmic management of workers through digital applications used for assigning, optimising, and evaluating human jobs through algorithms [6] for transport (Uber, Lyft, Ola), food delivery (UberEATS, DoorDash), and other forms of on-demand service. Despite the shift towards automation in the labour economy, platform labour continues to replicate precarious traditional work structures. In addition, digital work is burdened with lack of regulation and self-determination of workers. So called "independent contractors" are not covered by labour law or institutional, benefits [7], and are not eligible for labour protection, insurance, social security unemployment benefits, minimum wages, paid leave, severance packages, pensions, and/or retirement savings. Lack of geographical boundaries and decentralization of labour have encumbered the regulation of digital labour [8].

Digital work in the Global South

Dados defines the "Global South" as "regions outside Europe and North America, including Latin America, Asia, Africa, and Oceania that are either low-income and/or politically and culturally marginalized" [9]. In the Global South alone, there are an estimated 40 million platform workers. The digital space in Africa, Asia, and Latin America has seen a massive growth in the number of digital platforms and digital work both in local and international contexts. According to the report by the Oxford Internet Institute [10], Asia has witnessed a growing internet and mobile connectivity such that "India, Bangladesh, and Pakistan are now among the largest suppliers of gig workers to major digital platforms in the world" [11]. Although the adoption of digital labour is limited in Africa owing to minimal access to the Internet among populations [12], the continent has shown significant promise in creating opportunities for work online.

Digital Work as Emancipatory in the Global South

The process of digitalization has impacted the lives of digital labourers in the Global South [13] in myriad positive ways. As Aguilar et al. argue, the Global South market has created an "opportunity for informal workers to become formal" [14], and enabled greater flexibility and financial independence, particularly for women who can continue to engage in caregiving duties while working [15]. Platform labour has had a net positive effect on developing countries by reviving stagnant economies and creating job opportunities for daily wage workers. As digital work can be conducted remotely with Internet access, gig work has become borderless providing high levels of flexibility for workers with health reasons and/or domestic and care-giving responsibilities [16]. For women in the Global South, the gig economy has been a blessing in disguise. According to the Asia Foundation's research conducted on female gig workers, women find freedom in remote work by avoiding difficult workplaces, harassment from colleagues, and exhaustive schedules. Many women in the Global South become primary breadwinners for their families, and retain financial independence and agency in decisions of labour.

Furthermore, for daily wage workers who struggle with precarious work conditions, the stability of platform work and the supporting infrastructure with steady hourly wages and monthly income, digital labour can be a welcoming and rewarding alternative. Calvão describes digital work as a liberating and emancipatory experience for workers in the Global South, with workers enjoying free reign in regulation, anonymity, and lowered cost in algorithmic management [17]. However, despite its emancipatory potential, digital work's lack of regulation and self-determination of workers, particularly in the Global South, creates significant tensions and contradictions. There are few existing regulations or standards that seek to challenge concentrated power in the hands of stakeholders.

**Worker rights in the Global South**

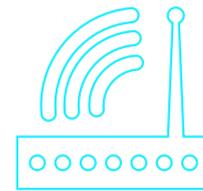

Lack of regulation and self-determination particularly affects the Global South; in countries in Asia and Africa where digital work is increasingly mediated; where labour is unbound by regulation, "prone to exclusion, sub-contracting and decreased bargaining power" [18]; and where it's affected by gender gaps and surveillance technologies. Although digital labour provides stable wages and income for gig workers, workers in the Global South experience discrimination either in relation to transnational contexts based on nationality, or in local contexts based on gender, race, and/or religion. Despite the common misconception that gig workers in the Global South are always un/underemployed daily wage workers, many are, in fact, well-educated [19] and well-versed in English.

However, owing to discrimination, some tasks do not reach those in the South [20]. Furthermore, there is a significant difference in pay in similar work held by workers in the North and South due to the difference in the minimum hourly wages. Wages are set and regulated by platforms without structured regulation. Aguilar et al. [21] argue that there is lack of clarity and transparency on "issues such as fair labour practices, access to benefits, taxation, social security, and dispute resolution leading to inferior labour outcomes" [22] for women and other marginalized groups. According to Berg et al. [23] and the ILO survey in 2017, there is evidence of lack of social protection, health insurance, pensions, retirement plans, and government assistance among surveyed participants. In addition, geographic distance and isolation has equally contributed to an inability to organize and self-collectivize.

Without formal contracts, the hired gig workers rarely have worker rights that come with full-time employment.

**Global South case study – Ola Cabs**

India is no exception to the rise of the gig economy where workers find labour on platforms for food delivery applications such as Zomato and Swiggy, transport applications such as Ola and Uber, data entry in Flipkart, and Amazon, and other forms of work. The gig economy has thrived in the country, and has engaged working professionals, daily wage workers, temporary workers, part-time workers, and workers from marginalized groups. Here, I explore the Indian cab hail company Ola to investigate the Indian gig model within the broader context of the Global South, worker rights, and minoritized groups involved with the platform.

Ola Cabs is an Indian ridesharing firm developed by ANI Technologies Private Limited. Primarily based in Bengaluru, Karnataka, India, the company offers services in major Indian cities. Established in 2010, Ola remains one of the largest mobility platforms in India, extending to around 20 million customers [24]. Ola's mobile application, similar to Uber, allows customers to connect with drivers nearby in order to book rides. Along with Uber, Ola has a captured market of 95% in the platform taxis business [25]. The company's business model relies on its digital platform that makes it easier for both drivers and passengers to use. In addition to allowing
drivers to use their own cars, Ola, in collaboration with financing  partners and car manufacturers, also leases cars to those that don't own them, thereby enabling drivers to work under the Ola banner.

Ola's digital platform is part of the massive gig economy in India that attracts labour for short-term gigs for Indian workers. The company guarantees drivers flexible hours, working schedules, and the freedom of work. However, drivers for Ola work as "independent contractors," and are therefore not covered by labour laws, and institutional benefits. They are not eligible for basic protection against risks associated with work, insurance, financial stability, unemployment benefits, minimum wages, paid leave, severance packages, pensions, retirement savings and/or social security benefits such as maternity benefits, disability cover, employment injuries [26]. Instead, drivers must work long hours during the day in order to pay their daily lease to the company that acts as a third-party ride hailing company. Undertaking full-time work does not equate to full-time employment for employees [27]. Therefore, Ola has complete monopoly and autonomy over worker rights, and forces Indian drivers and gig workers to adopt full time employment albeit without the benefits that they are contractually obligated to provide [28]. In addition to the lack of benefits and basic protection, workers take on further outsourced costs for vehicle maintenance, insurance and fuel. In order to finance the loans for car ownership, drivers borrow from banks or lease cars from Ola leading to larger debt traps and more worker exploitation [29]. Many drivers who work for Ola are tractor drivers migrating from rural villages owing to "unstable and low incomes, droughts, and low agricultural output," and are forced to put up their agricultural land and properties as collateral for car loans [30].

In India, Ola drivers are also subject to discrimination based on religion, caste, gender, sexual orientation, and socio-economic status. Many hailing from vulnerable religious and lower-caste minorities who face stigma in their villages also migrate to cities for more stable work and in search of better economic status and socio-economic mobility [31]. However, drivers are unable to remain anonymous, and undergo discrimination based on their religion and caste. Similarly, drivers themselves discriminate by declining booking from passengers, and leaving low ratings based on race, gender, and religion. In both cases, drivers receive low ratings that furthers their risk of employment within the precarious gig economy.

Further, Ola employs surveillance systems to track driver routes, the number of rides, time taken for each ride, and the passenger-acceptance rates. With complete access to the personal information of workers, Ola also monitors the navigation routes of workers, and their precise locations throughout the day, and controls workers through penalties [32]. Surveillance technologies coupled with paying systems employed by platform companies for gig work, therefore, engage in Foucault's biopolitics [33] where the employer has undue control over the work lives of gig workers in the precarious economy.

The COVID-19 lockdown has further exacerbated these risks, and drivers have lost substantial household income while they continue to pay towards leasing and renting their cars. Workers have recently demanded an extension on the Reserve Bank of India's moratorium on loan repayments, and continue to strike against high commission rates charged by Ola. In addition, gig workers are working towards the creation of a collective union to help alleviate the risks of surveillance, financial liabilities, and lack of social security benefits, particularly for vulnerable communities in India.

How can we address current problems?

Digital self-determination – worker rights, privacy, and autonomy

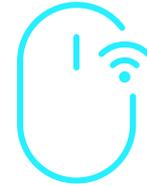

As observed in the Ola case study, there is a serious lack of transparency and clarity behind the functioning of platform algorithms that empower some while disempowering others. As the worker does not understand what goes on behind the algorithm, what decisions are being made, how their work is being evaluated by the algorithm, and how the data they generate is employed, they find it difficult to make proper decisions as to their conduct [34]. Owing to this lack of transparency, there emerges a power vacuum and imbalance that directly affects the worker, reducing their ability to switch jobs. In addition to issues in transparency, platforms also exercise significant control over the workers through constant surveillance, monitoring, and tracking of location [35]. Whether in the delivery business, or in the field of transport, workers are constantly monitored, with their locations, personal, and other information stored as data points for later reference. This in turn suggests an absence of privacy for workers who undergo surveillance even when outside the job, or while taking breaks. In this datafied economy of digital labour, where all worker interactions and data are collected, the data that is generated by the workers does not belong to them. This becomes even more pertinent in the Global South, where for the poor, the marginalized, and minoritized groups, surveillance of everyday transactions can become a matter of life or death. Digital self-determination as a concept is an emerging field of study that nvestigates questions of privacy, surveillance, and autonomy of digital citizens. In the context of platform workers, particularly in the Global South, self-determination is the enabling of a process of regaining worker data and privacy through an act of decolonization. By building worker trust through the design of reliable, user-centered platforms, and the decentralization of digital spaces where information is not held in the hands of a single corporation in the platform economy, gig workers can achieve more transparency, autonomy, and agency over the data they generate.

Regulation of digital work

Berg et al. [36] argue that the platform economy lacks the infrastructure or proper government regulation. Platforms ultimately have the power to decide how their workers, their contracts, and rights are regulated. Therefore, platforms have the ability to disempower collective action, and the construction of unions [37]. As digital labour is currently poorly integrated within regulatory and governmental frameworks, it is imperative that platforms integrate both universal regulatory frameworks for corporations, and provide rights for union building. This is even more pertinent owing to differing governmental laws, and legal frameworks in countries in the North and South, and owing to the geographic isolation of workers that disrupts the formation of collective unions, and effectively removes bargaining power from the workers. Regulatory frameworks must, therefore, be established and structured between the platform, the regulator, and workers as stakeholders [38]. This chapter has highlighted the larger tensions between the emancipatory potential and the lack of regulation and worker autonomy in the South through case studies in India.

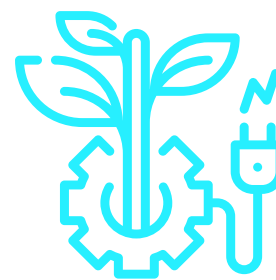

# 06
# MEDIA & COMMUNICATIONS & ETHICAL FORESIGHT

Matthew Hutson, Lujain Ibrahim, Heather von Stackelberg, Victoria Martín del Campo

Introduction

People who write and speak publicly about AI have a responsibility to know something about the field so that they can portray it accurately. They should also be aware of the narratives they're creating; stories that feed into fear or hype or visions of utopias or dystopias often sell well, but aren't good for society. They can also drown out more important matters. Portrayals of AI in journalism and elsewhere also need to include the broader context of the product, such as the source of the physical materials used, and what damage that extraction might be causing to people and the environment, as well as exploitative labour used for things like data labeling. Coverage of AI can also draw lessons from how writers cover wars, in that war reporters can add fuel to conflict, and potentially draw a lot of readers by being deliberately provocative; they can be as neutral as possible, even though neutrality is not truly possible; or they can focus on amplifying diverse voices and on conflict resolution. We suggest that coverage of AI should take this last approach, modeled after peace reporting.

Covering AI Responsibly

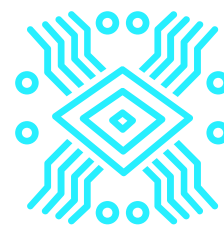

Artificial intelligence is a highly technical field that has a lot of jargon that can be difficult to understand, even for those working in the field. But it's important that the general public understand the technology as well as they can, as it impacts their lives directly and indirectly every day. AI feeds into virtual assistants like Siri and Alexa, search engines, customer tracking and ad displays, fraud detection, driving navigation, product and movie recommendations, social media and news feeds, facial recognition, credit ratings, parole decisions, and medical diagnoses and prognoses. The average person may not have a hand in programming the computers or deciding which algorithms to deploy, but they can make their voices heard through protests, social media posts, communication with legislators, product and service purchases, and software user settings.

The average person may not have a hand in programming the computers or deciding which algorithms to deploy, but they can make their voices heard through protests, social media posts, communication with legislators, product and service purchases, and software user settings.

A majority of people get their information about AI from movies, TV shows, and social media [1], which are not always accurate. Other sources need to fill that information gap. These sources include news media such as newspapers, magazines, websites, and TV, as well as public information officers at universities and companies, scientists and engineers, entrepreneurs, advertisers, bloggers, public intellectuals, politicians, activists, and anyone else with an audience, online or off.

Now is an especially exciting time to cover AI, as the advances pile up quicker and quicker and as they seep into more and more areas of our lives—and tomorrow will be an ever more exciting time. But lest anyone get too caught up in the excitement, communicators should follow certain guidelines [2] to avoid painting utopian or dystopian pictures, and audiences should ask certain questions about the information they receive.

First, communicators and audiences need a basic understanding of the field. A few key ideas are worth touching on frequently. First, the definition of AI is amorphous and contested. Some simple algorithms receive the label to wow and amaze, while some forms of automation are downplayed because they're threatening or not that smart. Second, AI comes in at least three flavors: machine learning, or ML, which gets all the press these days; other optimization algorithms, and Good Old-Fashioned AI, or GOFAI. The first relies on lots of data to find patterns and make predictions. The second searches the space of possible solutions. The third relies on manually programmed decision rules. Finally, AI is very narrow. People can do lots of things well, but each AI program, or "model" (a common term for machine learning software), can do only one or a few things well. We are not close to AGI—artificial general intelligence.

It's easy to overpromise AI's capabilities, because we anthropomorphize it, attributing human abilities or characteristics. It can do some of the things we do, so we unconsciously or consciously assume it can do many of the other things we can do. But it lacks common sense. A program that performs well on one task might not even perform well on the same task in slightly different conditions. For instance, a program that diagnoses disease from X-rays may work on images from one X-ray machine but not another. AI systems are also vulnerable to "adversarial attacks," in which an attacker elicits nonsensical or entirely inappropriate answers. Changing a few pixels of an image can make software see, say, a photo of a school bus as an ostrich.

When discussing what AI can do, it's important to acknowledge what it can't do. In reporting a successful test of an algorithm, a reporter might want to mention failed tests. Or to place quotes around certain words like "believes" or "imagines" or "wants" to clearly demarcate them as metaphors. And researchers don't appreciate clickbaity headlines and images from The Terminator. When describing neural networks, a popular type of machine learning, it should be made clear that they only roughly mimic the biological neural networks in human brains. Even eminent scholars sometimes compare

The AI Ethics Brief   |   The MAIEI Learning Community Report

the number of parameters in a neural net to the number of synapses in a brain, as if algorithms will become as smart as brains when the two numbers match up. [3]

Journalists should take extra time to evaluate claims in press releases and blog posts, especially those on new products or services. What exactly is new, if anything? Is there anything the source might not want you to know? Ask if the source has any academic research to back up the claims made. The general public should also be skeptical of announcements that speak in generalities or technical-sounding buzz words, especially if the technology is all proprietary and not open to evaluation by outside experts. This is especially the case when developers—either in academia or in industry—talk about an AI system that does amazing things with really high accuracy. It's very easy to get very high accuracy answering questions in carefully set-up test environments that are entirely unlike what would exist in the "real world". When journalists parrot those accuracy statistics without questioning them or investigating the set-up, they contribute to the misunderstanding of AI and what it's capable of. [4]

Often researchers publish work on the preprint server arXiv ("archive") before it's peer-reviewed at a journal or conference. If covering this work, journalists should do their own peer review and ask outside experts to comment on it.

When interviewing researchers, journalists should ask detailed questions about how the work was done. What question or problem inspired the work? What challenges were overcome? What was learned? What's next? Going into more detail, if it's a machine-learning system, how was the training data collected and labeled, and how was the system trained and evaluated? How were comparison methods selected and prepared? Has the system been tested in real-world settings?

And of course, ask about applications and limitations. Researchers may highlight positive uses of their technologies, but it's also important to include possible abuses. A system that mimics human language might improve therapy chatbots, but it could also troll people online. Has the researcher thought about how we might avoid or ameliorate such harms? And even for positive uses, any technology has weaknesses or ways it could be improved. What are some ways in which we might not (yet) want to rely on this new system?

Coverage of AI will ideally include the voices not only of those who create or deploy it, but also those willingly or unwillingly affected by it. These could include customers or other users, or victims—for instance people targeted by surveillance. AI can benefit some people while harming others, so sources should come from diverse backgrounds. These perspectives can inform public policy or consumer decisions, and they might even guide the future development of technology. Women and people of color and those from lower socioeconomic backgrounds make up a disproportionately small part of the tech elite, so they may raise points or concerns that coders haven't fully considered.

When covering or reading about technology, we should keep in mind that technology is created by, and for, people.

### Responsibility for Hype

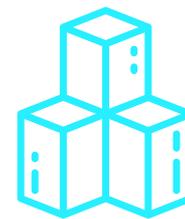

Noam Chomsky once said, "The smart way to keep people passive and obedient is to strictly limit the spectrum of acceptable opinion but allow very lively debate within that spectrum." Whether people are doing this deliberately or not, this seems to be a tendency of discussions of AI ethics. There is robust discussion of the trolley problem and which solutions are culturally acceptable, and where they're acceptable… but this is a not-very-realistic, entirely hypothetical scenario, relevant only for fully self-driving vehicles that are likely a decade or more away. This discussion crowds out discussions of issues and harms that are happening right here and now, like how AI systems are being used in justice contexts, and whether they should be used there at all.

But it's stories that get people's attention, especially good stories. And good stories always have conflict, and they usually have some sort of threat, danger, bad guy, or conspiracy. So in order to get people's attention, people talking in public, like the media, talk about how AI is going to take over everyone's jobs, or how AI is going to create a dystopian nightmare—or about how it's going to create a utopia. This feeds into fears and hype, and it becomes very difficult to counter these narratives with more realistic scenarios, because these fearful, exaggerated ones are such good stories. Nearly everyone recognizes the robot from The Terminator and knows that story; many fewer people know what robots can actually do right now.

There are a number of problems with that bias. One is that it allows companies that are doing ethically questionable things with AI to fly under the radar of public attention, because so much attention is being paid to making sure that AGI (which may or may not arrive this century) will be beneficial to human beings. For example, today, hiring and security companies are using AI emotion detection systems that are operating based on discredited scientific underlying theory and that may have very low accuracy for people from non-Western countries, yet the answers provided by these systems are being used for making high stakes decisions. This isn't a sexy story that will capture people's imaginations, but right now, it's directly impacting far more people. The spectrum of debate is far too limited.

All the fear, hype, and overblown hopes sell stories, but other people are taking advantage of it, too. Estimates are that around 40% of "AI-based" start-ups in Europe actually involve no AI whatsoever [5]—they generally are using humans (often low-paid humans) to achieve the effect, and just market themselves as using AI. They do this to cash in on the hype. Similarly, management consulting companies are using the hype (and contributing to it, too) in order to sell their services for "digital transformation" or an "AI transformation" to companies, playing on fears that the companies will be left behind and become obsolete if they don't. The management consulting companies often charge huge consulting fees for this service, despite the fact that they also often have very little technical expertise on their teams, as these companies generally recruit from business schools, not computer science programs.

Another problem with stories of fear and hype is the impact they have on public policy and on research funding. Public-policy makers often respond to the narratives that are resonating with the public. As a result, they can create unnecessary or overly restrictive policies that severely hamper the growth of an area, or they can ignore an area and not put necessary policies into place. Both of these can cause harm to the industry and to the people directly affected by the AI systems that are being poorly regulated. [6]

There have already been some AI "summers" in which there was a lot of hype—and a lot of research funding available as a result—and some AI "winters" in which disillusionment about AI's slow progress resulted in funding cuts and reduced enthusiasm for AI research. Narratives that capture the public's imagination, whether realistic or not, can strongly influence what gets funding, and what doesn't.

Instead of these narratives of fear and hype, we need both individuals and organizations who are willing to call out bad behavior, especially on the part of large companies that otherwise aren't very accountable for their actions.

This means, among other things, that whistleblowers need protection from backlash and abuse, such as what Timnit Gebru experienced from Google when she refused to retract criticism of large language models such as those Google uses. People speaking in public, such as journalists and bloggers, can call for public policies that protect people such as Dr. Gebru. Disagreeing with her criticisms and discussing them is one thing; abuse and death threats are something else—and Dr. Gebru got as much or more of the latter than she did the former. [7] Journalists and bloggers should also themselves make sure to not add to the abuse and disrespect.

Organizations are much better situated than individuals are to do the investigations and bring bad behavior to the public's attention. For example, most people did not realize how widely the COMPAS system was being used, and how biased it was against Black Americans, until the investigative journalism organization ProPublica published an in-depth investigative report. [8] Even though there's been some disagreement about the validity and rigor of ProPublica's approach, there's been no disagreement that Black Americans are not being treated fairly by the American justice system or the COMPAS system. [9] ProPublica's article, though it hasn't significantly changed the widespread use of the COMPAS system, has spurred a significant, ongoing discussion within the AI industry about fairness, and how it is defined and operationalized within AI systems.

This is what we need the news media to do—call out companies that are contributing to harm, either through bad policies or through ignorance or inattention to what their AI systems are doing. There aren't nearly enough organizations doing this.

A few decades ago, Big Tobacco strategically leveraged their PR machine and ability to fund researchers to prevent widespread public discussions about the health hazards of smoking. They were very much aided in this by the small number of media channels available at the time, which each had a very large audience. Big Tech is pulling from Big Tobacco's playbook in terms of providing funding for research, but their own products have removed the media advantage that was well used by tobacco companies. [10] Still, Dr. Gebru's experience shows that they have weapons that they won't hesitate to use, and use effectively. We need both individuals and organizations that are willing to do the investigative work and call out bad behavior. We need public discussions, and a critical mass of responsible journalists.

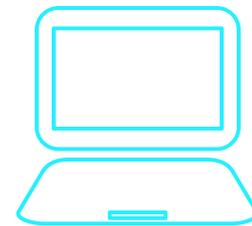

### Highlighting AI's "Hidden" Harms

In the case of AI systems, one way to expand "the spectrum" Chomsky refers to is to insist that the media increases its emphasis on the full "anatomy" of AI systems. In Anatomy of an AI System, a visual essay outlining the extraction and exploitation behind AI systems, the co-authors Vladan Joler and Kate Crawford quote Mark Graham on contemporary capitalism [11]:

> contemporary capitalism conceals the histories and geographies of most commodities from consumers. Consumers are usually only able to see commodities in the here and now of time and space, and rarely have any opportunities to gaze backwards through the chains of production in order to gain knowledge about the sites of production, transformation, and distribution.

Joler and Crawford's essay takes Amazon's Echo as a starting point and dissects it, as one of the many commonplace AI devices that is enabled by contemporary capitalism and runs on large amounts of material resources, consumer data, and human labor. They show that in the context of the AI industry, this is a prevalent reality at all levels of development, deployment, and discardment that is driven by profit maximization and inseparable from historical and present-day colonialism.

From extracting essential minerals in mines to manufacturing hardware to labelling training data, human and natural resources are at the center of the
63

The AI Ethics Brief   |   The MAIEI Learning Community Report

"extractive processes" that allow for the existence of a lot of digital infrastructure that enables AI systems. This exploitation of human labor and natural resources is characterized by "uneven geographies" and is concentrated in developing countries [12].

However, dominant communication and coverage of AI systems, their benefits, and their harms remains limited in framing and leaves out significant and consequential sections of the anatomy of these systems. In other words, it "conceals the histories and geographies" of commodities in the digital economy like data and hardware. Building mainstream awareness of the full anatomy of AI systems should be a central objective of communicating the harms and limitations of these systems.

Thus, defining the parameters of this discussion in a constricted fashion where algorithmic harm is presented as mostly resulting from biased datasets and mostly solvable by technological fixes is ahistorical and misleading at best, and at worst legitimizing of the harms resulting from the extraction and exploitation that precede and succeed algorithmic design. Greater public awareness of "the sites of production, transformation, and distribution" is needed to contest these practices, which marginalize and exploit. Expanding "the spectrum" also weakens "technological solutionism" narratives, as it exposes more of the human and environmental costs. As Joler and Crawford note, even the Echo user herself is "simultaneously a consumer, a resource, a worker, and a product."

### Scrutinizing Big Tech's AI Communications

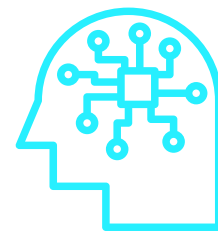

Another challenge facing the communication of AI and AI ethics is responding to Big Tech's increasingly delusive communication of their human-AI systems. In a humorous yet poignant critique of Big Tech's co-optation of "AI ethics,"

Karen Hao of MIT Technology Review presents the following definition of "accountability" taken right out of big tech's dictionary [13]:

> accountability (n) - The act of holding someone else responsible for the consequences when your AI system fails.

As Hao notes, with many of big tech's products, practices, and power being put under public and government scrutiny, companies have amped up their PR efforts to reframe discussions on the failures and shortcomings of their human-AI systems. A notable example of such sociotechnical systems is big tech's content moderation systems. This effort has also included the development of "ethics" and "oversight" boards, like the now-dissolved Google AI Ethics board and the Facebook Oversight Board, which critics believe is an effort to redefine what "accountability" can look like and preserve the status quo of unchecked power [14]. While the Google AI Ethics board was quickly dissolved due to controversies over its board members and lack of real power to address concerns with Google's AI systems, Facebook's Oversight Board, which is composed of 20 members from different countries and backgrounds, remains functional and has been running since October of 2020 [15, 16].

Facebook Oversight Board's main function is to make precedent-setting content moderation decisions, where a large portion of media coverage of censorship, Facebook, and the Facebook Oversight Board has recently concentrated on the board's decision to uphold the account suspension of former U.S. President Donald Trump.

However, content moderation on Facebook and other platforms also happens at a much larger scale and uses a combination of automated moderation and human moderation. Deploying AI systems for content moderation has been repeatedly promised, especially by Facebook CEO Mark Zuckerberg, as the solution [17] to detecting and removing hate speech, misinformation, fake accounts, and more. However, experts have expressed doubts over the maturity of the technology, and Facebook's failures to moderate content appropriately have already led to disastrous consequences (including but not limited to fueling the genocide in Myanmar) [18]. So how has Facebook responded to criticism over its content moderation and censorship?

A recent censorship effort by Facebook that has garnered media attention is the transnational repression of Palestinian and pro-Palestinian content on Facebook's platforms. Activists, as well as organizations like the Palestinian digital rights organization 7amleh, have documented these acts of censorship for years [19,20]. This censorship has recently reached an unprecedented scale following the increased documentation and sharing of state and settler-inflicted displacement and violence against Palestinians in May of 2021, with companies like Facebook blocking and restricting millions of posts and accounts [21]. One example of the automated aspect of this censorship was reported by BuzzFeed News, which found that Instagram was censoring posts that use the hashtag of Islam's holiest mosque (Al Aqsa Mosque), located in Palestine [22].

In response to the increased condemnation regarding these acts, Facebook fended off criticism by continuing its strategy of swinging the accountability pendulum between blaming "technical glitches" and blaming "human moderators/errors" [23]. Resorting to the convenience of blaming automated systems and technical errors is not new to the company. According to The Verge, when Mark Zuckerberg testified before Congress in 2018, he cited AI as the solution to harm (from unmoderated hate speech to bias) Facebook is creating over 30 times [24]. This strategy effectively allows Facebook to escape accountability and further questioning due to the catchall nature of the term "AI," and the lack of transparency over what the nature and scale of Facebook's human-AI collaborations look like. These collaborations come in different forms, from raumatized human moderators at Facebook to underpaid workers annotating data [25]. Many of these workers may introduce their own unconscious biases and/or may be pushed into majority thinking by the requesters or the crowdsourcing platforms themselves [26].

What this shows is the need for the media to continually push back and scrutinize big tech companies' PR efforts to deflect criticism and accountability in the form of co-opting "AI ethics" language as well as muddling the details of their human-AI efforts. This is also necessary to build the public's AI literacy into one that is more resilient to illegitimate claims of ethical development and governance, especially as big tech companies continue to avoid any real checks on their power.

Lessons from War Reporting

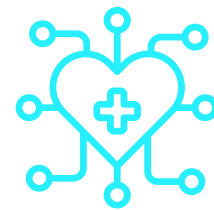

An obvious conclusion from all of this is that the media have become today the main source of information about Artificial Intelligence (AI). Consequently, they exert a fundamental influence on the audience's understanding of that disruptive technology and, indirectly, on their attitude towards it. Moreover, such is the relevance of the "media universe" that we can affirm that it affects not only the way technology is perceived, but also its economic, cultural and political functioning. Thus, the media's actions in a given social context are not innocuous; the media are not mere observers of social processes but participants in them.

This role is especially relevant in the specific case of technological misuse and conflicts, where the news media inevitably participates in the conflict cycle. The media acts simultaneously as a sources of information, innovation, weapons, targets, and challenges, and it has an important role in the creation of social attitudes in relation to technology and their transformation; thus, the media can encourage aggressive behavior, justify unfair tech abuse actions, orm stereotypes, images of the enemy and demonizations. Perhaps the best known of the many documented examples, as shown in the paper Fanning the Flames of Hate: Social Media and Hate Crime [27], is the Anti-refugee campaigns in Germany's case, a paragon of the so-called "hate media," although examples are abundant, as shown by research on the media's influence on technological propaganda. However, the media can also arouse compassion, mercy and charity; stimulate demands for action or policy changes; or break distances.

Thus, we can find examples in which the media have put into practice their ability to denounce human rights violations or to reduce the level of tension in conflict situations and promote peace processes. Among these trends, the implementation of peace journalism stands out, a relatively recent paradigm (its origin usually dates back to the 1970s) that is rooted in peace research and is based on the application of conflict theory and analysis to the field of communication. Precisely, the aim of this section is to gather the main characteristics of what has been called peace journalism, whose essential features distinguish it from the currently dominant journalistic paradigm, highlighting its practical, analytical and theoretical dimensions and contributions, as well as the barriers and obstacles that hinder its development and application today.

Conflict, violence, and, in short, bad news constitute the raw material par excellence of the information offered by the media. In the context of wars and armed conflicts, this preference seems to be accentuated. In the 1960s, one of the promoters of peace journalism, the Norwegian researcher Johan Galtung, reached the same conclusions. Galtung and Ruge (1965), based on the analysis of the coverage of a series of conflicts in four Norwegian newspapers, suggest: first, that acts of violence (here technological misuse) become newsworthy events in themselves and, second, that when dealing with the issue of violence, the media always overlook one factor: peace. Galtung (1998a) suggests that here are two journalistic models or paradigms when dealing with conflicts: the dominant or war journalism and the alternative or peace journalism First, the hegemonic paradigm or war journalism has the following features (Galtung, 2002):

- Bad journalism is strongly influenced by propaganda. WState, war journalism tends to identify with one side or, in any case, with the defense of national interests. It also tends to systematically privilege the dominant political and economic forces. This can be translated to the tech world as making the tech career a national conflict. The common cartoon of countries against each other such as this news title: "China and the US are locked in a superpower tech war to 'win the 21st century,'" ignores the complexities of those countries, their governments and corporations and their real role in tech.
- Bad journalism is based on the opinion of the elites, who become the main source of information. The media, in the search for a supposed objectivity, turn to sources considered prestigious, i.e., the elites and official sources, which, thus, end up influencing the media agenda.
- In bad journalism, conflicts are simplified, reducing the parties involved to a few sides framed in a process in which "what one side wins is what the other loses." In tech, this false dichotomy can be represented by portraying "tech vs. humans" or "corporations vs. governments" and other types of cartoons.
- In bad journalism, conflicts are simplified, reducing the parties involved to a few sides framed in a process in which "what one side wins is what the other loses." In tech, this false dichotomy can be represented by portraying "tech vs. humans" or "corporations vs. governments" and other types of cartoons.

Tech journalism assumes the traditional criteria that make a fact newsworthy, basing its contents on concrete events, limited in time, decontextualized, recent and easily explainable. In this way, it is direct physical violence that is the focus of their interest, while other effects that are not so reproducible in images are disregarded. The result, regardless of the initial intention of the professionals, is that this form of journalism contributes to the conflict, to the creation of irreconcilable camps and to hide the peace processes and proposals that may be developing.

In contrast to the dominant practice, Galtung proposes the paradigm of peace journalism, oriented towards conflict transformation, which attends to the voice of the victims and interprets peace as a process in which all parties must be involved and receive benefits. Journalists who fit into this perspective report on existing peace initiatives, provide information that contextualizes conflicts and consider their non-directly visible effects, as well as their structural and cultural causes.

In short, it is journalism involved in the processes of conflict resolution, reconstruction and reconciliation, which highlights the shared elements among the actors and sectors involved and not only the differences. Peace journalism is thus a challenge to war journalism, in this case violence and misuse of technology, insofar as it focuses on contextualization and long-term processes and, especially, on the diversification of topics and sources of information, actively seeking those voices that represent options for peaceful conflict resolution. The following section details, to a greater extent, the characteristics and dimensions that make up the alternative paradigm of peace journalism.

| Peace Journalism | Conflict Journalism |
| --- | --- |
| - Explores conflict formation, actors, objectives, issues, win/win perspective<br>- Gives an open space, open time; causes and links everywhere, also in history and culture<br>- Makes conflict transparent without turning any conflict into violence<br>- Gives voice to all parties<br>- Focus on the invisible effects of violence: trauma, structural and cultural damage, etc. | - Attention focused on the conflict scenario, two parties, one objective (win, victory)<br>- Closed space, closed time; causes and exits on stage, who threw the first stone<br>- Making conflicts opaque, secret- "Us-them", propaganda, voice to the us- "Them" as the problem, who predominates<br>- Reactive: waiting for violence and closes communications channels |

In the recent past of AI reporting, there has been a tendency to focus on how evil, greedy corporations are going to harm everyone, or control them, or steal everything, or how amazing, innovative corporations are going to save us all. There's little attention paid to the data sources, the data work, and the physical inputs necessary for a technical product, with more focus on who is at fault than on how it could be repaired and the same thing avoided in the future. In contrast, writers could look at why people are in conflict with corporations, and what perspectives are missing where. They could give voice to those who usually aren't heard by AI developers, and show the damage that is done —- usually unintentionally—by poor design.

Conclusion

For AI to advance in an ethical way, the media needs to cover AI responsibly, and they also need to cover "Responsible AI" responsibly — the two go hand in hand. Covering AI responsibly includes being realistic (avoiding hype and clickbait) about the merits and limits of these systems, asking informed and critical questions not only about the algorithm's performance but also about how it was developed and whom it will affect, and playing an active role in increasing the public's "AI literacy." As responsible coverage of AI continues to reveal the social implications and limits of these systems, "Responsible AI" is often presented as the solution to these limitations. However, to avoid problem simplification, ethics washing, and tech solutionism, covering "Responsible AI" responsibly and rigorously must also occupy a critical space in responsible AI media coverage. A framework for achieving this can be drawn from peace reporting, as contrasted with war reporting, emphasizing conflict resolution and diverse voices and sources.

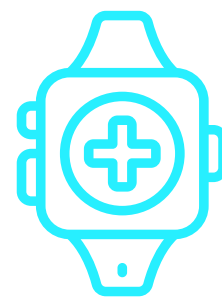

# 07

# CONCLUSION

I hope that was as much of a pleasure reading for you as it was for me!

Upon reflecting on the report, I find myself enriched and enchanted by the information that has been presented to us. Our aim was to generate a space both multitalented filled with a wide range of expertise, a space which the AI realm desperately needs more of. Whether it is to serve as an example as to how technology can be both the "medicine" and the "poison", or how we have a responsibility to report AI responsibly, the benefits of these efforts are clear.

What will you take away from this report? We are always delighted to hear from you and incorporate your suggestions in making this report a more valuable resource for the entire community here at MAIEI. So, please don't hesitate in reaching out to us at support@montrealethics.ai to chat and share your ideas.

In the meanwhile, and until your eyes once again set upon our pages, stay safe and take care of those around you. For, as expressed in my personal favourite saying from the philosophy of Ubuntu:

I am because we are and since we are, therefore, I am.

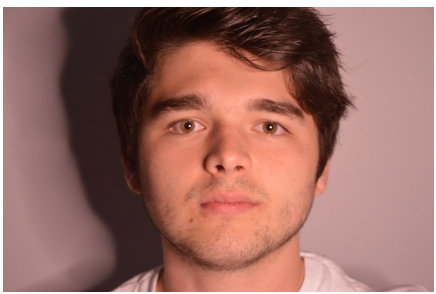

### Connor Wright, Partnerships Manager

[Connor Wright](#)

[@Csi_wright](#)

# 08 SUPPORT OUR WORK

The Montreal AI Ethics Institute is committed to democratizing AI Ethics literacy. But we can't do it alone.

Every dollar you donate helps us pay for our staff and tech stack, which make everything we do possible. With your support, we'll be able to:

- Run more events and create more content.
- Use software that respects our readers' data privacy.
- Build the most engaged AI Ethics community in the world.

Please make a donation today at montrealethics.ai/donate.

We also encourage you to sign up for our weekly newsletter The AI Ethics Brief at brief.montrealethics.ai to keep up with our latest work, including summaries of the latest research & reporting, as well as our upcoming events. If you want to revisit previous editions of the report to catch up, head over to montrealethics.ai/state.

Please also reach out to Masa Sweidan masa@montrealethics.ai for providing your organizational support for upcoming quarterly editions of the State of AI Ethics Report.

Note: All donations made to the Montreal AI Ethics Institute (MAIEI) are subject to our Contributions Policy.

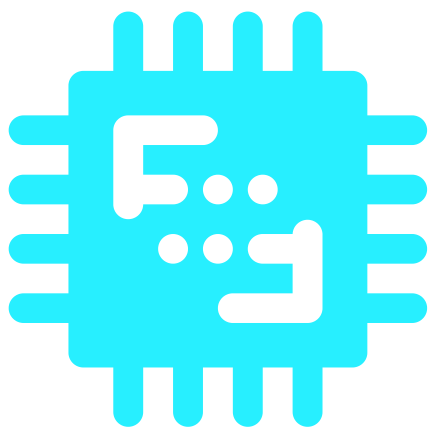

# REFERENCES AND FOOTNOTES

## Chapter 3

[1] https://www.theverge.com/2019/2/25/18229714/cognizant-facebook-content-moderator-interviews-trauma-working-conditions-arizona

[2] On the Opportunities and Risks of Foundation Models
https://arxiv.org/pdf/2108.07258.pdf

[3] The Race to Understand the Exhilarating, Dangerous World of Language AI
https://www.technologyreview.com/2021/05/20/1025135/ai-large-language-models-bigscience-project/

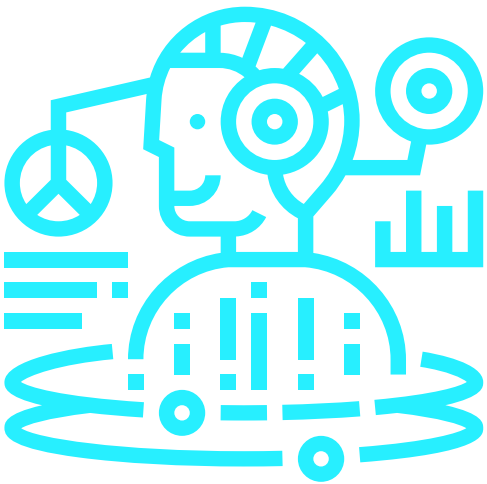

# Chapter 4

[47] United Nations High Commissioner for Refugees. (n.d.). Freedom on the Net 2012 - Burma. Retrieved from https://www.refworld.org/docid/5062e8a81e.html

[48] Ibid.

[49] Ibid.

[50] Tønnesson, S., Oo, M. Z., & Aung, N. L. (2021). Pretending to be States: The Use of Facebook by Armed Groups in Myanmar. Journal of Contemporary Asia, 1-26. doi:10.1080/00472336.2021.1905865

[51] United Nations High Commissioner for Refugees. (n.d.). Freedom on the Net 2012 - Burma. Retrieved from https://www.refworld.org/docid/5062e8a81e.html

[52] Whitten-Woodring, J., Kleinberg, M. S., Thawnghmung, A., & Thitsar, M. T. (2020). Poison If You Don't Know How to Use It: Facebook, Democracy, and Human Rights in Myanmar. The International Journal of Press/Politics, 25(3), 407-425. doi:10.1177/1940161220919666

[53] United Nations High Commissioner for Refugees. (n.d.). Freedom on the Net 2012 - Burma. Retrieved from https://www.refworld.org/docid/5062e8a81e.html

[54] Tønnesson, S., Oo, M. Z., & Aung, N. L. (2021). Pretending to be States: The Use of Facebook by Armed Groups in Myanmar. Journal of Contemporary Asia, 1-26. doi:10.1080/00472336.2021.1905865

[55] Whitten-Woodring, J., Kleinberg, M. S., Thawnghmung, A., & Thitsar, M. T. (2020). Poison If You Don't Know How to Use It: Facebook, Democracy, and Human Rights in Myanmar. The International Journal of Press/Politics, 25(3), 407-425. doi:10.1177/1940161220919666
80

The AI Ethics Brief  |  The MAIEI Learning Community Report

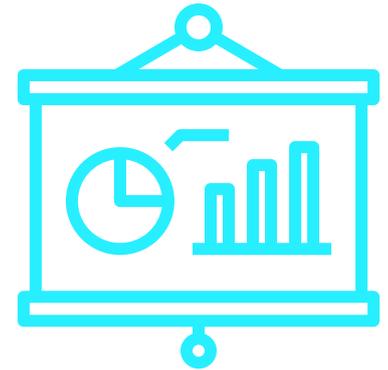

[121] Isaac, M., & Goel, V. (2020, April 22). Facebook Invests $5.7 Billion in Indian Internet Giant Jio. The New York Times. https://www.nytimes.com/2020/04/21/technology/facebook-jio-india.html

[122] Facebook Invests $5.7 Billion in INdia's Jio Platforms. (2020, April 22).

[123] TechKee Admin. (2020, April 21). Facebook invests $5.7B in India's Reliance Jio. TechKee. https://www.techkee.com/facebook-invests-5-7b-in-indias-reliance-jio/

[124] Isaac, M., & Goel, V. (2020, April 22). Facebook Invests $5.7 Billion in Indian Internet Giant Jio. The New York Times. https://www.nytimes.com/2020/04/21/technology/facebook-jio-india.html

[125] Mukherjee, R. (2018). Jio sparks Disruption 2.0: infrastructural imaginaries and platform ecosystems in "Digital India." Media, Culture & Society, 41(2), 175–195. https://doi.org/10.1177/0163443718818383

[126] Staff. (2017, September 5). 10 Ways the Indian Telecom Industry Changed After Jio Started Operations. NDTV Gadgets 360. https://gadgets.ndtv.com/telecom/features/jio-10-ways-the-indian-telecom-industry-changed-after-jio-started-operations-1746342

[127] Mukherjee, R. (2018). Jio sparks Disruption 2.0: infrastructural imaginaries and platform ecosystems in "Digital India." Media, Culture & Society, 41(2), 175–195. https://doi.org/10.1177/0163443718818383

[128] TechKee Admin. (2020, April 21). Facebook invests $5.7B in India's Reliance Jio. TechKee. https://www.techkee.com/facebook-invests-5-7b-in-indias-reliance-jio/

[129] Mukherjee, R. (2018). Jio sparks Disruption 2.0: infrastructural imaginaries and platform ecosystems in "Digital India." Media, Culture & Society, 41(2), 175–195. https://doi.org/10.1177/0163443718818383

[130] Ibid.
88

The AI Ethics Brief   |   The MAIEI Learning Community Report

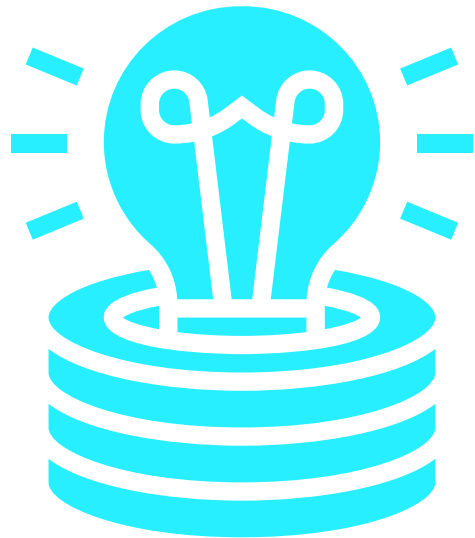

# Chapter 5

# Chapter 6

27] Müller, Karsten and Schwarz. 2020. Fanning the Flames of Hate: Social Media and Hate Crime. Princeton University. https://papers.ssrn.com/sol3/papers.cfm?abstract_id=3082972

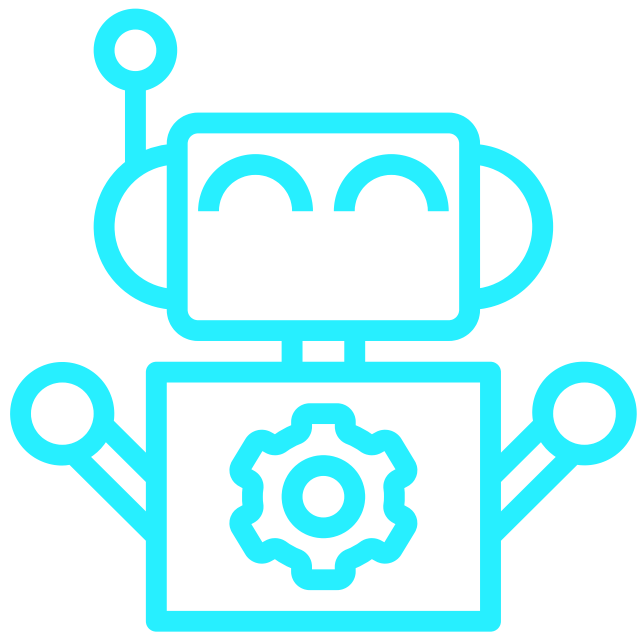